\documentclass[]{aastex631}
\shorttitle{Value added catalog for LAMOST DR8}
\shortauthors{Wang C. et al.}
\graphicspath{{./}{figures/}}

\begin{document}

\title{The value-added catalogue for LAMOST DR8 low-resolution spectra}

\author{Chun Wang}
\altaffiliation{LAMOST FELLOW}
\altaffiliation{Corresponding author (wchun@tjnu.edu.cn)}
\affiliation{Tianjin Astrophysics Center, Tianjin Normal University, Tianjin 300387, People's Republic of China.}
\affiliation{Department of Astronomy, Peking University, Beijing 100871, People's Republic of China.}
\affiliation{Kavli Institute for Astronomy and Astrophysics, Peking University, Beijing 100871, People's Republic of China.}

\author{Yang Huang}
\affiliation{South-Western Institute for Astronomy Research, Yunnan University, Kunming, Yunnan 650091, People’s Republic of China.}

\author{Haibo Yuan}
\affiliation{Department of Astronomy, Beijing Normal University, Beijing 100875, People’s Republic of China.}

\author{Huawei Zhang}
\affiliation{Department of Astronomy, Peking University, Beijing 100871, People's Republic of China.}
\affiliation{Kavli Institute for Astronomy and Astrophysics, Peking University, Beijing 100871, People's Republic of China.}

\author{Maosheng Xiang}
\affiliation{Max-Planck Institute for Astronomy, Konigstuhl, D-69117, Heidelberg, Germany.}

\author{Xiaowei Liu}
\affiliation{South-Western Institute for Astronomy Research, Yunnan University, Kunming, Yunnan 650091, People’s Republic of China.}



\begin{abstract}

We present a value-added catalog containing stellar parameters estimated from 7.10 million low-resolution spectra for 5.16 million unique stars with spectral signal-to-noise ratios (SNRs) higher than 10 obtained by the Large Sky Area Multi-Object Fibre Spectroscopic Telescope (LAMOST) Galactic spectroscopic surveys. 
The catalog presents values of stellar atmospheric parameters (effective temperature $T_{\mathrm{eff}}$, surface gravity $\log g$, metallicity [Fe/H]/[M/H]), $\alpha$-element  to metal abundance ratio [$\alpha$/M], carbon and nitrogen to iron abundance ratios  [C/Fe] and [N/Fe] and 14  bands'  absolute magnitudes  deduced from LAMOST spectra using the method  of neural network. The spectro-photometric distances of those stars are also provided  based on the distance modulus.  For stars with spectral SNRs larger than 50, precisions of   $T_{\mathrm{eff}}$,  $\log g$,  [Fe/H], [M/H], [C/Fe], [N/Fe] and [$\alpha$/M] are 85\,K, 0.098\,dex, 0.05\,dex, 0.05\,dex, 0.052\,dex, 0.082\,dex and 0.027\,dex, respectively.  The errors of 14  band's absolute magnitudes are 0.16--0.22\,mag for stars with spectral SNRs larger than 50.   The spectro-photometric distance is accurate to 8.5\% for  stars with spectral SNRs larger than 50, and is more accurate than geometrical distance for stars with distance larger than 2.0\,kpc.  Our estimates of [Fe/H] are reliable down to  [Fe/H] $\sim -3.5$\,dex, significantly better than previous results. The catalog provide 26,868 unique very metal poor star candidates ([Fe/H] $\leq -2.0$). 
The catalog would be a valuable data set to study the structure and evolution of the Galaxy, especially the solar-neighbourhood and the outer disc.

\end{abstract}



\section{Introduction} \label{sec:intro}

As a typical disc galaxy,   the Milky Way (MW) is an ideal  laboratory to test  galaxy formation and  evolution scenarios with the advantage that the individual  stars in the MW can be well resolved.  
A number of completed or ongoing large-scale spectroscopic surveys, e.g., RAVE \citep{Steinmetz2006}, SEGUE \citep{yanny-segue}, LAMOST Experiment for Galactic Understanding and Exploration \citep[LEGUE; ][]{deng-legue,liu-lss-gac, Zhao2012}, Gaia-ESO \citep{Gilmore2012}, Galactic Archaeology with HERMES \citep[GALAH; ][]{DeSilva2015}, Apache Point Observatory Galactic Evolution Experiment \citep[APOGEE; ][]{Majewski2017}, Gaia Radial Velocity Spectrometer \citep{Cropper2018}, and the MMT H3 survey \citep{Conroy2019} — as well as the upcoming surveys such as SDSS-V \citep{Kollmeier2017}, 4MOST \citep{Feltzing2018, deJong2019}, and WEAVE \citep{Dalton2014}, have propelled studies of  the structure, stellar populations and the chemical and dynamic evolution of the MW. 
Stellar atmospheric parameters and elemental abundances of stars could be accurately derived from high-to-low resolution spectra collected by those massive spectroscopic surveys.
Those rich information from spectroscopic surveys, together with the  accurate parallax and proper motions from the Gaia missions \citep{gaiadr1,gaiadr2, Gaiaedr3}, will significantly advance our knowledge of the formation and evolution of our Galaxy.

The LAMOST Galactic survey \citep{deng-legue,liu-lss-gac, Zhao2012} is the first spectroscopic survey to obtain spectra of $\sim$10 million stars. 
For the LAMOST spectra, several efforts have been made to derive the radial velocities $V_{r}$,  stellar atmospheric parameters (effective temperature $T_{\mathrm{eff}}$, surface gravity $\log\,g$, metallicity $ \mathrm{ \, [Fe/H]} $), individual element abundances, distance and extinctions \citep{yuan2015a, lamost-dr1, LSP3, Xiang2017}.  
The official LAMOST Stellar parameter Pipeline \citep[LASP; ][]{lamost-dr1} and the LAMOST Stellar Parameter Pipeline at Peking University \citep[LSP3; ][]{LSP3, Xiang2017} have been developed to estimate the radial velocities and  atmospheric parameters from LAMOST spectra  for A/F/G/K-type stars, with typical uncertainties of  $T_{\mathrm{eff}}$, $\log\,g$, $ \mathrm{ \, [Fe/H]} $ and $V_{r}$, respectively,  of 100--200\,K, 0.1--0.3\,dex, 0.1--0.2\,dex and 5 $\rm km\,  s^{-1}$  \citep{Gao2015, lamost-dr1, Ren2016, LSP3, Xiang2017,Wang2016,Huang2018} for FGK type stars with spectral SNR\,$>\,10$.
$\alpha$-element to iron abundance ratio [$\alpha$/Fe] and abundances of individual elements (C, N, O, Na, Mg, Al, Si, Ca, Ti, Cr, Mn, Fe, Co, Ni, Cu, and Ba) are also derived \citep{Liji2016, Xiang2017, Xiang2019, Ting2017}, for SNR\,$>$\,50, their uncertainties are 0.03--0.1\,dex except Cu and Ba (with uncertainties of 0.2--0.3\,dex).
In addition, some other efforts have been made to derive stellar parameters from LAMOST spectra, e.g.,  the Stellar LAbel Machine \citep[SLAM; ][]{Zhang2020}, the analysis based on SP\_Ace code   \citep{Boeche2018}, the stellar parameter determinations based on the $Cannon$   \citep{Ho2017}, an application of the SEGUE Stellar Parameter Pipeline (SSPP) to LAMOST spectra   \citep{Lee2015} and the hotPayne from deriving stellar labels for hot, OBA stars \citep{Xianghot}.   Value-added catalogs of stellar parameters from these efforts have been widely utilized for unraveling structure  and evolution of our Galaxy \citep[e.g.,][]{ Huang2015, Sun2015,Xiang2015, Chen2016, Huang2016, Liu2017, Xiang2018, Xu2018, Huang2019, Kamdar2019, Gandhi2019,  Li2019, Wang2019a,Wang2019,Sharma2021b,Sharma2021,Vickers2021}.

Due to the lack of very metal-poor stars (VMP; [Fe/H] $\leq -2.0$) as the spectral templates or training samples, the stellar labels are generally less well estimated for very metal-poor stars with [Fe/H] $<-2.5$.
The [Fe/H] values of VMP stars need to be estimated separately by special considerations \citep{Li2018}.  
The PASTEL  \citep{Soubiran2010} catalog provides accurate stellar atmospheric parameters including $T_{\mathrm{eff}}$, $\log\,g$ and [Fe/H]. 
Moreover, the PASTEL catalog contains tens of VMP stars (in common with LAMOST) with accurate stellar atmospheric parameters, which can be adopted as the training set for LAMOST to estimate the stellar atmospheric parameters, especially the metallicity values of VMP stars.
For the estimates of [$\alpha$/Fe] and individual elemental abundance  [X/H] from LAMOST low-resolution spectra, previous works mostly use the APOGEE\,DR15/DR14 or GALAH as training set for giant and dwarf stars.
Compared with APOGEE\,DR15/DR14,  the APOGEE DR16 provides more precise measurements of metallicity [Fe/H], $\alpha$-element to metal/iron abundance ratio [$\alpha$/Fe]/[$\alpha$/M], individual elemental abundance  [X/H]  and individual elemental abundance to iron ratios [X/Fe], especially  for dwarf stars.  Adopting the APOGEE\,DR16 as training set, we could obtain good estimates of  individual elemental abundances for both giant and dwarf stars.

The absolute magnitudes present the luminosity of stars and thus are useful for   estimating the distances and ages of stars, which are indispensable when we study the structures and chemo-dynamical evolutions  of our Galaxy.  
The  early Gaia DR3 \citep[hereafter EDR3;][]{Gaiaedr3} have  provided  parallaxes, proper motions of 1.8 billion Galactic stars.  
The geometrical distances  \citep{Baileredr3} of 1.47 billion stars in the Milky Way have also been provided.  
Within 2\,kpc from the Sun, the geometrical  distance  is accurate to 10\,per\,cent.  
The uncertainties and systematic errors of geometrical  distance  become larger for distant stars, especially those beyond 2\,kpc. 
For those stars far from ($> 2$\,kpc) the Sun, the spectro-photometric distance is a better choice, as it is less distance-dependent and may provide better precision \citep[e.g., ][]{Xiang2017, Xiang2021}. 
The common stars of LAMOST, Gaia EDR3 and other photometric surveys provide a good training set to estimate  absolute magnitudes  from spectra directly for  deriving the sepctro-photometric distances for all stars with LAMOST low-resolution spectra. 

By March 2021,  Low-Resolution Spectroscopic Survey of LAMOST DR8 have released 11,212,561 optical (3700--9000\,\AA) spectra with a resolution of $R$$\sim$1800, of which more than 90 per\,cent  are stellar spectra. The classifications, radial velocity $V_{r}$ and stellar atmospheric  parameters  for those spectra are provided by LAMOST\,DR8.  

In this work, we  present a  value-added catalog for  LAMOST\,DR8 low-resolution spectra. 
The catalog  provide stellar atmospheric parameters ($T_{\mathrm{eff}}$, $\log\,g$,  $ \mathrm{ \, [Fe/H]} $/[M/H]),  absolute magnitudes of 14 different wavelength bands adopted by different photometric surveys ($G, Bp, Rp$ of Gaia EDR3, $J, H, K_{\rm s}$ of 2MASS, $W1, W2$ of WISE, $B, V, r$ of APASS and $g, r, i$ of SDSS),  elemental abundances to metal or iron ratios ([$\alpha$/M], [C/Fe], [N/Fe]) and spectro-photometric distances for stars targeted by LAMOST with their low-resolution spectra in LAMOST\,DR8.  We firstly  exploite a machine learning method based on neural network (NN)  to estimate effective temperature, surface gravity, metallicity, absolute magnitudes, and individual elemental abundances from the LAMOST\,DR8 low-resolution spectra, utilizing the aforementioned spectral training sets.  The distances of those stars are then estimated using the apparent magnitudes, extinctions and our estimated absolute magnitudes.  

The paper is organized as follows.  Section\,2 introduces the used neural network method and models. Section\,3 presents the training sets. In Section\,4, we present the estimates of stellar parameters  derived from  LAMOST spectra using neural network method, including a detailed error analysis.  We present the improved metallicity estimates of very metal poor stars in Section\,5. 
In Section\,6, we introduce the estimates  of spectro-photometric distance.  
We discuss the nature of the final value-added catalog  and present a perspective of the catalog on the future in Section\,7, followed by a summary in Section\,8.

\section{The neural network method and models}

$T_{\mathrm{eff}}$, $\log\,g$,  $ \mathrm{ \, [Fe/H]} $ and other chemical elemental abundances are  stellar parameters that can be derived from stellar spectra.  
As discussed by  \cite{Xiang2020}, the luminosity (absolute magnitude) of a star is related to the effective temperature,  mass and surface gravity  of the star according to the Stefan-Boltzmann equation and gravity equation.  
The stellar  mass also  implicitly depends on $T_{\mathrm{eff}}$, $\log\,g$, $ \mathrm{ \, [X/H]} $ and stellar rotation.  
Thus, the luminosity (absolute magnitude) could be directly derived from stellar spectra.   
In this section, we introduce neural network method and models, which model the empirical relation between the LAMOST spectra and 
$T_{\mathrm{eff}}$, $\log g$, $ \mathrm{ \, [Fe/H]}$, [M/H], [$\alpha$/M], [C/Fe], [N/Fe] and absolute magnitudes of 14 photometric bands aforementioned using different training sets.

\subsection{The neural network models}
To build up a relation between  spectra and effective temperature, surface gravity, element abundances and absolute magnitudes of stars, we adopt a feed-forward multi-layer perceptron neural network model, which is similar to that of  \cite{Ting2017, Ting2019} and \cite{Xiang2020,Xiang2021}. Adopting the Einstein sum notation, our neural network contains three-layers and can be succinctly written as:
\begin{equation}
P=\omega\sigma(\omega^{'}_{i}\sigma(\omega^{''}_{j}\sigma(\omega_{\lambda_{k}}f_{\lambda}+b_{k})+b_{j})+b^{'}_{i}),
\end{equation}
 where $P$ is  the stellar atmospheric parameters, element abundances or absolute magnitudes of stars;  $\sigma$ is the Relu activation function; $\omega$ and $b$ are
weights and biases of the network to be optimized; the index
$i$, $j$ and $k$ denotes the  number of  neurons in  the third, second and first layer; and $\lambda$ denotes the wavelength pixel. The total number of neurons for first, second and third layer are respectively  512, 256 and 64.  The training process is carried out with the $Tensorflow$ package in $Python$. 
 
\subsection{Pre-processing for the spectra}
 The LAMOST spectra  cover a  wavelength range of $\lambda\lambda$3700--9000. However, we only use 3900-5800\,\AA\,and 8450--8950\,\AA\,to determine the  effective temperature, surface gravity, element abundance ratios and absolute magnitudes of stars.  First the SNR of  very blue part (3700--3900\,\AA) of the spectra is quite low for most of stars.  Secondly, the spectra of 5800--8450\,\AA\,suffer from serious background contamination (including sky emission lines and telluric bands). Besides, the effective information in spectra of 5800--8450\,\AA\, is  very limited. 
 
Because of the unknown extinction values for most of the survey targets and  the relatively large uncertainties of spectral flux calibration of LAMOST spectra,  it is better to use continuum-normalized spectra to deliver reliable stellar parameters  \citep [e.g.,][]{Xiang2017}.  All LAMOST stellar spectra are normalized  by dividing their pseudo-continuum spectra.  Each  pseudo-continuum spectrum is derived through using a sixth-order polynomial to fit  the smoothed spectra of wavelength 3900-5800\,\AA\, and  8450--8950\,\AA\,, separately. The smoothed spectra are  obtained via  median filtering method with a filter size of 21 pixels. 
 
For  neural-network method,  to avoid the issues due to the different scales in different dimensions of the input data, all the normalized spectral fluxes and the stellar labels of the training sets should be normalized in a standard space (with zero mean and unity variance).  All LAMOST normalized spectra are also normalized in standard space with the same scale of training sets.

\section{The training  sets}
In the current work, we define three training sets to estimate stellar paramters: the LAMOST--PASTEL common stars, the LAMOST--APOGEE common stars and the LAMOST--Gaia\,EDR3--2MASS--WISE--APASS--SDSS\,(hereafter referred to as ``the LGMWAS")\,common stars.  The LAMOST--PASTEL common stars are used to estimate effective temperature $T_{\mathrm{eff}}$, surface gravity $\log g$ and metallicity  $ \mathrm{ \, [Fe/H]} $. The LAMOST--APOGEE common stars are used to estimate the  $\log g$, [Fe/H], [M/H], [$\alpha$/M], [C/Fe] and [N/Fe]. Note that here we will give two $\log g$ and [Fe/H] values for each low-resolution spectra in LAMOST\,DR8, which  are estimated using  LAMOST--PASTEL common stars and  LAMOST--APOGEE common stars as training sets, respectively.  The LGMWAS common stars are used to estimate absolute magnitudes ($M_{G},M_{Bp}, M_{Rp}$ of Gaia EDR3 bands, $M_{J}, M_{H}, M_{Ks}$ of 2MASS bands, $M_{W1}, M_{W2}$ of WISE bands, $M_{B}, M_{V}, M_{r_{A}}$ of APASS bands, $M_{g}, M_{r}, M_{i}$ of SDSS bands).  Tabel\,\ref{table1} presents the  three adopted training sets and their effective parameter ranges.


\begin{deluxetable}{cccc}
\tablecaption{The three adopted training sets,  estimated stellar parameters, number of stars and effective parameter ranges using those three adopted training sets.\label{table1}}
\tablehead{
\colhead{Training sets} & \colhead{Parameters} & \colhead{Number of stars} &\colhead{Effective parameter range}
}
\startdata
& & & $3\,500 < T_{\mathrm{eff}} < 10\,000\,$K  \\
 LAMOST--PASTEL &  $T_{\mathrm{eff}}$, $\log g$, [Fe/H]   &1,281  &  $0 < \log g < 5$\\
  & & &  $-3.8 < [\rm Fe/H] < 0.5$\\
  \hline
& & & $3\,800 < T_{\mathrm{eff}} < 6\,500\,$K \\
 & & &  $0 < \log g < 5$\\
 & & &  $-2.3 < [\rm Fe/H] < 0.5$\\
  LAMOST--APOGEE &$\log g$, [Fe/H], [M/H], [$\alpha$/M],  [C/Fe], [N/Fe] &8,753  &  $-2.3 < [\rm M/H] < 0.5$\\
  & & &  $-0.2 < [\alpha/\rm M] < 0.4$\\
  & & &  $-0.6 < [\rm C/Fe] < 0.4$\\
  & & &  $-0.3 < [\rm N/Fe] < 0.6$\\
\hline
 & & &  $-2.5 < M_{G} < 7.5$ \\
  & & &  $-2 < M_{Bp} < 8$ \\
   & & &  $-2.5 < M_{Rp} < 7$ \\
  & & &  $-4 < M_{J} < 6$ \\
   & & &  $-5 < M_{H} < 5$ \\
   & & &  $-5 < M_{Ks} < 5$ \\
LGMWAS & $M_{G},M_{Bp}, M_{Rp}$, $M_{J}, M_{H}, M_{Ks}$, $M_{W1}, M_{W2}$, $M_{B}, M_{V}, M_{r_{A}}$, $M_{g}, M_{r}, M_{i}$ &6,000&    $-5 < M_{W1} < 5$ \\
   & & &  $-5 < M_{W2} < 5$ \\
   & & &  $-1 < M_{B} < 8$ \\
    & & &  $-2 < M_{V} < 8$ \\
   & & &  $-2 < M_{r_{A}} < 8$ \\
   & & &  $-1 < M_{g} < 8$ \\
    & & &  $-2 < M_{r} < 8$ \\
   & & &  $-2 < M_{i} < 8$ \\
\enddata
\end{deluxetable}

\subsection{The LAMOST--PASTEL common stars}

PASTEL is  a catalog of bibliographical compilation of accurate stellar atmospheric parameters, including $T_{\mathrm{eff}}$, $\log g$ and [Fe/H], mostly determined from high resolution, high SNR spectra.  
By 2020, the catalog  contains 81,362 stellar atmospheric parameter measurements for more than 31,000 unique stars, collected from more than 11,000 bibliographical references \citep{Soubiran2016, Soubiran2020}.  
However,  stellar atmospheric parameters coming from different  bibliographical references do not have the same scales.  
Not all of them are mutually independent, systematic deviations among different bibliographic sources may exist because of  the different approaches of measuring stellar atmospheric parameters. Besides, not all values of effective temperature and surface gravity listed in PASTEL were determined with high resolution spectra.  
To solve those problems in PASTEL, we have recalibrated the measurements of stellar atmospheric parameters collected in PASTEL  from a variety of sources such that they are on the same scale. The results will be presented in a separate paper (Huang et al. 2021, in preparation). 

\begin{figure*}
\centering
\includegraphics[width=5.5in]{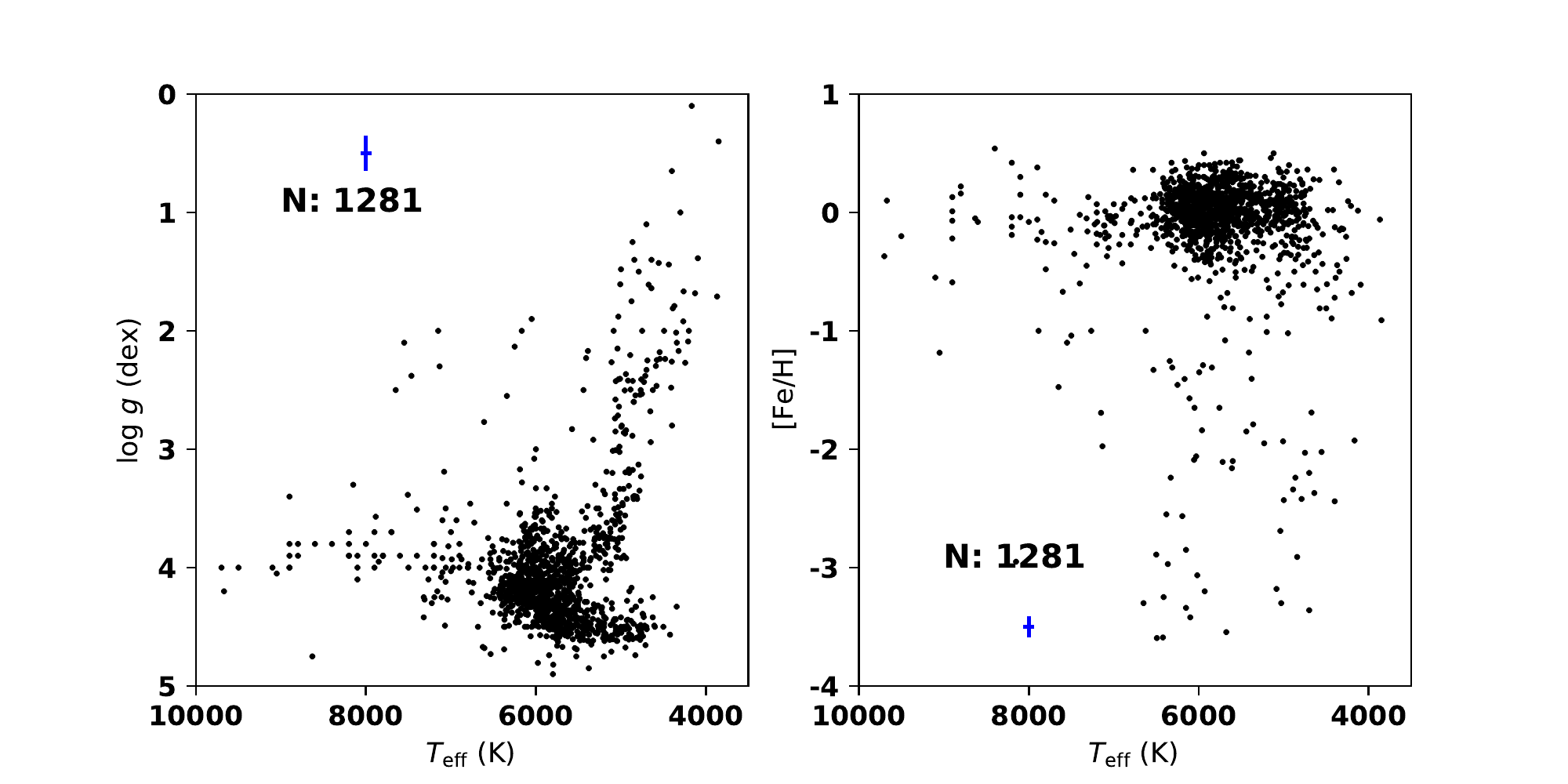}
\caption{The distributions of the 1,281 LAMOST--PASTEL training stars in the $T_{\mathrm{eff}}$--$\log\,g$  (left panel) and $T_{\mathrm{eff}}$--[Fe/H] (right panel) planes.  Typical errors of $T_{\mathrm{eff}}$, $\log\,g$ and [Fe/H] of the LAMOST--PASTEL training sample are indicated by the  error bars (blue cross) in each panel. }
\label{parameter_space}
\end{figure*}

We find 1,859 common unique stars with good  stellar atmospheric parameter determinations through a cross-match between  recalibrated PASTEL and LAMOST\,DR8. 
Amongst them, 1,743 unique stars have good LAMOST spectra (without bad pixels in the adopted wavelength range, check via eyes),  1,281 unique stars have high quality  LAMOST spectra with spectral SNR\,$>$\,50.  
These 1,281 unique stars are finally adopted as training set. Other 462 uniqe stars  are adopted as test set.  
Fig.\,\ref{parameter_space} shows the distributions of the 1,281 LAMOST--PASTEL training stars in the $T_{\mathrm{eff}}$--$\log\,g$   and $T_{\mathrm{eff}}$--[Fe/H] planes. 
With this training set, we estimate the $T_{\mathrm{eff}}$, $\log g$ and  [Fe/H].  
The training set has a wide coverage of stellar atmospheric parameters: $3,500 < T_{\mathrm{eff}} < 10,000$\,K, $0.0 < \rm{log}$$\,g < 5.0$\,dex and $-4.0 < $[Fe/H]$ < 0.5$\,dex. 

With the LAMOST--PASTEL training set, we apply neural-network models to construct the relations between the LAMOST spectra and the stellar atmospheric parameters.  
We use the neural-network models  to estimate $T_{\mathrm{eff}}$, $\log\,g$  and  [Fe/H] for  LAMOST--PASTEL training and test sets.  
The estimated $T_{\mathrm{eff}}$, $\log\,g$  and  [Fe/H] could match well with that provided by the PASTEL catalog.  The standard deviations of their differences are 75\,K for $T_{\mathrm{eff}}$,  0.09\,dex for $\log g$ and 0.05 dex for [Fe/H] for the training set, 
156\,K for $T_{\mathrm{eff}}$, 0.29\,dex for $\log g$ and  0.15 dex for [Fe/H] for the test set. 
The standard deviations of test set are much larger than that of training set. 
This may be the consequence of that the test sample has lower LAMOST spectral SNRs ($20 <$ SNR $< 50$)  and much less stellar members compared to the training sample. 
The systematic differences between stellar atmospheric  parameters coming from  PASTEL and the stellar atmospheric parameters derived from LAMOST spectra are very small with $\Delta {T_{\mathrm{eff}}} \sim $10\,K, $\Delta {\log g} < $0.01\,dex and $\Delta {\rm [Fe/H]} <$\,0.01\,dex for both training and test sets.   One can see Fig.\,\ref{pastel_test} in Appendix for more details. 

It is surprising that the  [Fe/H] values derived from LAMOST spectra has no obvious systematic errors  down to [Fe/H] $\sim -2.2$ (Fig.\,\ref{pastel_test} in Appendix) through comparing with that of PASTEL, which is a big improvement for [Fe/H] determinations from LAMOST low-resolution spectra. It is important for us to build up catalog of VMP stars  and carry out follow-up observations of those VMP stars found from the LAMOST.

\subsection{The LAMOST--APOGEE common stars}

APOGEE DR16 \citep{apogeedr16} contains 473,307  high-resolution ($R \sim$\,22,500), multiplexed, near-infrared (15140--16940\,\AA) spectra for about 437,445  unique stars covering both the northern and southern sky. These spectra are  collected using the APOGEE-N (north) instrument \citep{Wilson2019} in combination with the 2.5\,m Sloan Foundation telescope \citep{Gunn2006} at Apache Point Observatory (APO) in New Mexico and the APOGEE-S (south) with the  2.5 m du Pont telescope \citep{Bowen1973} at Las Campanas Observatory in Chile (``LCO 2.5\,m").  
Adopting these spectra,  \cite{apogeedr16} estimate accurate  stellar atmospheric  parameters ($T_{\mathrm{eff}}$, $\log\,g$ and [Fe/H]/[M/H]), $\alpha$-element to metal abundance ratio [$\alpha$/M], individual elemental (C, C\,{\sc {i}},  N, O, Na, Mg, Al, Si, P, S, K, Ca, Ti, Ti\,{\sc{ii}}, V, Cr, Mn, Fe, Co, Ni, Cu, Ge, Rb, Ce, Nd, and Yb) abundance ratios [X/H], and carbon and nitrogen  to iron ratios [C/Fe] and [N/Fe] based on ASPCAP \citep{Garcia2016}. These stellar atmospheric parameters and these chemical elemental abundances are well estimated and calibrated.   The typical uncertainties of $T_{\mathrm{eff}}$, $\log\,g$ and individual elemental abundance are respectively 98\,K,  $< 0.1$\,dex and $< 0.08$\,dex, one can see tabel\,6, tabel\,7 and tabel\,8 of \cite{apogeedr16} for details. 

In this work, we  adopt the  APOGEE DR16 as the training set to estimate the $\log g$, [Fe/H], [M/H], [$\alpha$/M], [C/Fe] and [N/Fe].  As discussed in Section\,2, we use the LAMOST spectra of 3900-5800\,\AA\,and 8450-8950\,\AA. The selected wavelength range contains spectral features of C, N and $\alpha$-elements (e.g., CN\,$\lambda4215$\AA, CH\,$\lambda$4314\AA, Ca\,{\sc{i}}\,$\lambda$4226\AA\, and Mg\,b lines etc.).  Theoretically,  [$\alpha$/M], [C/Fe] and [N/Fe] could be derived from LAMOST spectra. Previous works  \cite[e.g.,][]{Liji2016, Xiang2017, Xiang2019, Ting2017,Ho2017, Zhang2020} have also made efforts to determine the carbon, nitrogen and $\alpha$-element abundance using LAMOST low-resolution spectra and support that even low-resolution spectra could be used to estimate stellar individual element abundances. 
 
\begin{figure*}
\centering
\includegraphics[width=7.5in]{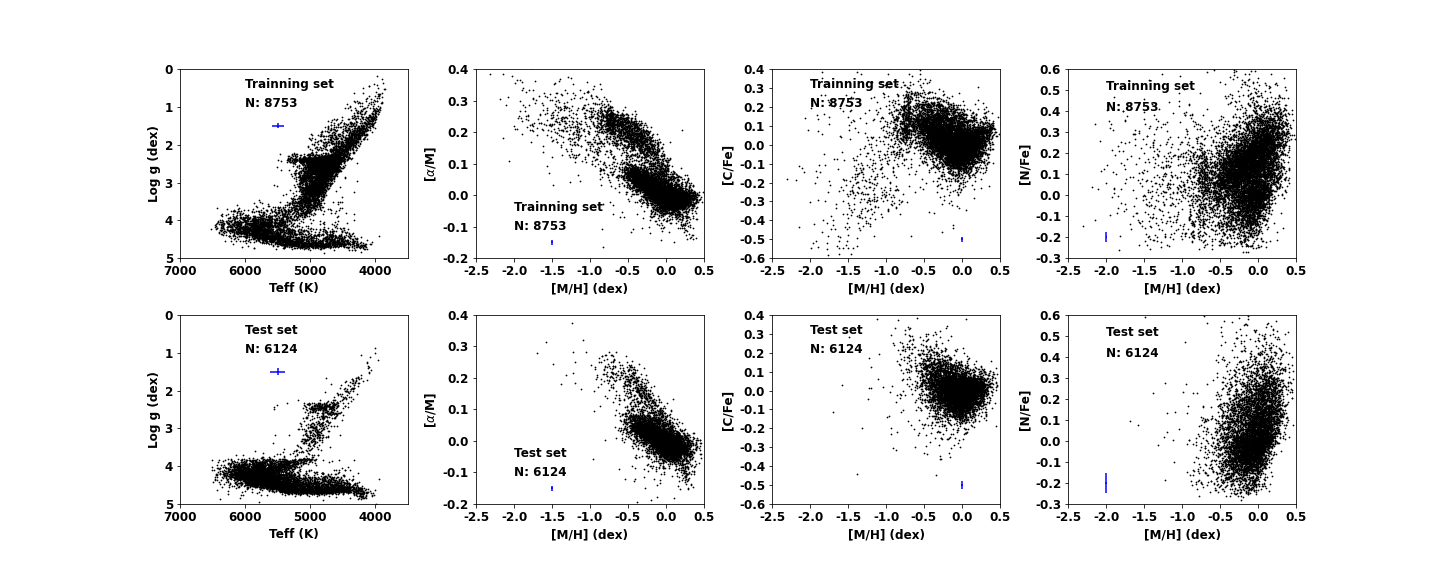}
\caption{The distributions of the LAMOST--APOGEE training stars (top panels) and test stars (bottom panels) in the $T_{\mathrm{eff}}$--$\log\,g$,  [M/H]--[$\alpha$/M], [M/H]--[C/Fe] and [M/H]--[N/Fe] planes. The number of stars in the training and test sets are labeled in each panel. Typical errors of $T_{\mathrm{eff}}$, $\log\,g$, [M/H], [C/Fe] and [N/Fe] of the LAMOST--APOGEE training sample are indicated by the  error bars (blue cross) in each panel. }
\label{parameter_apogee}
\end{figure*}

 A cross-match of LAMOST DR8 with the APOGEE DR16 catalog yields 122,658 stars in common. To derive accurate results, we select stars with the  following criteria: $\rm SNR_{LAMOST} > 80, SNR_{APOGEE} > 70,  ASPCAPFLAG \neq7, ASPCAPFLAG \neq 23, Err_{[Fe/H]} < 0.1, Err_{\emph{T}_{eff}}< 150\,K,\,  Err_{log\,\emph{g}} < 0.15,  Err_{[\alpha/M]} < 0.03, Err_{[N/Fe]} < 0.2, Err_{[C/Fe]} < 0.2$.   Stars with bad LAMOST spectra  are also excluded. Finally, 14,877 stars are selected.  We select 8,753 stars as training set, 6,124 stars as test  set. In this process, we select as much  metal-poor  and high [$\alpha$/Fe] stars as possible into  training sample in order to construct a better neural network model.   Fig.\,\ref{parameter_apogee} shows the distributions of the LAMOST--APOGEE training stars and  test stars in the $T_{\mathrm{eff}}$--$\log\,g$, [M/H]--[$\alpha$/M], [M/H]--[C/Fe] and [M/H]--[N/Fe] planes.    The training set has a wide coverage of stellar atmospheric parameters: $3,800 < T_{\mathrm{eff}} < 6,500$\,K, $0.0 < \log\,g < 5.0$\,dex, $-2.3 < $[M/H]$ < 0.5$\,dex, $\rm -0.2 < [\alpha/M] < 0.4$\,dex, $\rm -0.6 < [C/Fe] < 0.4\,dex$ and $\rm -0.3 < [N/Fe] < 0.6\,dex$. 

Note that when we estimate the  [C/Fe], we remove stars with $\rm Err_{[C/Fe]} > 0.05$. When we estimate the  [N/Fe], we remove stars with $\rm Err_{[N/Fe]} > 0.08$.  The purpose of the cuts is to  derive accurate [C/Fe] and [N/Fe]. 

Using the aforementioned neural network models adopting the LAMOST--APOGEE  training stars  as training set,  the relations between LAMOST spectra and aforementioned stellar parameters are constructed. 
Then we estimate the [Fe/H], [M/H], [$\alpha$/M], [C/Fe], [N/Fe] and $\log g$ \,for test stars using these models. 
The estimated stellar parameters could match well with that provided by APOGEE.
 The standard deviations of the residuals are only 0.033, 0.032,  0.044, 0.07, 0.018 and 0.067\,dex for  [Fe/H], [M/H],  [C/Fe], [N/Fe], [$\alpha$/M] and $\log g$, respectively.  The systematic differences between APOGEE chemical abundances and chemical abundances derived from LAMOST spectra are very small, almost all of them are smaller than 0.01\,dex.  The systematic differences between APOGEE $\log g$ and $\log g$ derived from LAMOST spectra is 0.012\,dex. 
 One can see Fig.\,\ref{test_apogee} in Appendix for more details about the comparisons between our estimated stellar parameters  and those provided by  APOGEE DR16. 

\subsection{The LGMWAS common stars }
Gaia EDR3 \citep{Gaiaedr3} provided precise parallaxes, proper motions and magnitudes of 1.8 billion stars in the MW.  
\cite{Baileredr3} estimate the distances and asymmetric uncertainties  for  1.47 billion stars by using Gaia EDR3.  
Within 2\,kpc from the Sun, the distance is accurate to 10\,per\,cent.  According to the precise distances, apparent magnitudes and interstellar extinctions  of stars, we can estimate the luminosity (absolute magnitudes) for these stars based on the distance modulus. The interstellar extinction of all LAMOST stars are estimated using the ``star-pair" method \citep{ Yuan2013}. 

 The LAMOST\,DR8, Gaia\,EDR3 and photometric surveys of 2MASS, WISE,  APASS, SDSS have more than 4 millions common stars.  
 We can take those stars with high quality data in spectroscopy, astrometry and photometry as training set to directly estimate  stellar absolute magnitudes from the LAMOST spectra.  
 The uncertainties of absolute magnitudes of  stars in the training set, which  depend on the accuracy of Gaia distances, apparent magnitudes, should smaller than 0.3\,mag.  LAMOST spectra of stars in the training set should have good quality. 
Thus, we select training stars with the following criteria: distance provided by Gaia\,EDR3  is smaller than 2\,kpc, the relative error of  distance are smaller than 15\,per\,cent, uncertainties of $G, Bp, Rp$ band magnitudes from Gaia\,EDR3 are smaller than   0.01\,mag,  uncertainties of $J, H, K_{s}$ band magnitudes from 2MASS are smaller than   0.03\,mag, uncertainties of $W1, W2$ band magnitudes from WISE are smaller than   0.035\,mag, uncertainties of $B, V, r$ band magnitudes from APASS are smaller than   0.035\,mag, uncertainties of $g, r, i$ band magnitudes from SDSS are smaller than   0.01\,mag and the SNR of LAMOST spectra is larger than 100. In order to minimize the effect on the estimated absolute magnitude of interstellar extinction corrections, we exclude stars with the Galactic latitude $|b| < 30^{\circ}$ and stars with interstellar extinction $E(B-V) > 0.02$\,mag.  
Though  the interstellar extinction  of most of the training stars are negligible, we have corrected for the extinctions using the values estimated with the ``star-pair" method \citep{ Yuan2013}.  Stars with bad LAMOST spectra are also removed from the sample.  Finally, we obtain 9\,086 stars.  Amongst them, 6,000 stars form the training set, 3,086 stars construct the test set. 

\begin{figure*}
\centering
\includegraphics[width=7.5in]{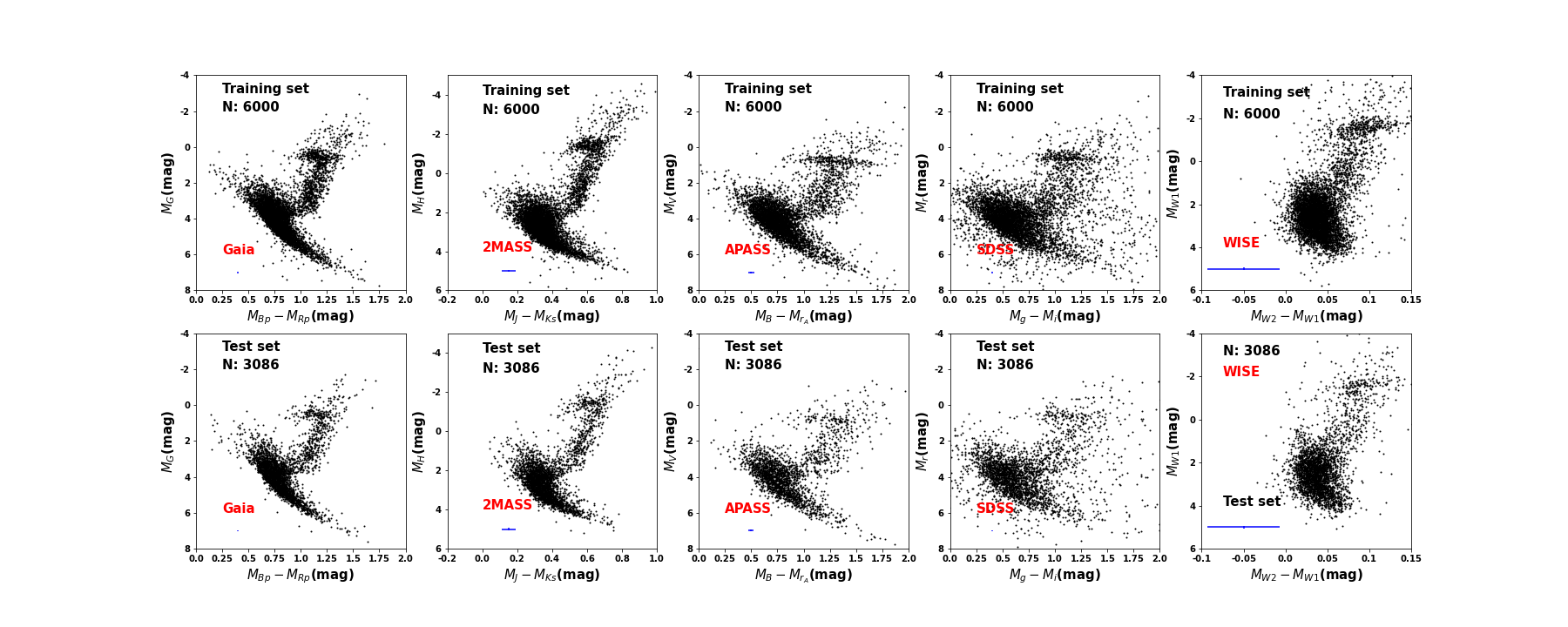}
\caption{The distributions of the LGMWAS  training stars (top panels) and test stars (bottom panels) in the ($M_{Bp}-M_{Rp}$)--$M_{G}$,  ($M_{J}-M_{Ks}$)--$M_{H}$, ($M_{B}-M_{r_{A}}$)--$M_{V}$ (APASS bands), ($M_{g}-M_{i}$)--$M_{r}$ (SDSS bands) and ($M_{W1}-M_{W2}$)--$M_{W1}$ planes. The number of stars in the training and test sets are labeled in each panel. Typical errors of these absolute magnitudes and colors of  the LGMWAS training sample are indicated by the  error bars (blue cross) in each panel.  The error bar in the y-axis is hard to see  because the error of absolute magnitude is much smaller than its coverage. }
\label{hr_diagram1}
\end{figure*}

Fig.\,\ref{hr_diagram1}  shows the distributions of LGMWAS  training stars and test stars in the Hertzsprung-Russell diagram plotted with Gaia\,EDR3 $G, Bp, Rp$, 2MASS $J, H, Ks$, APASS $B, V, r$, SDSS $g, r, i$ and WISE $W1, W2$ bands.  The training and test set have wide coverages  in the Hertzsprung-Russell diagrams.   
The two samples cover many types of stars, e.g., main-sequence stars, red giant stars and red clump stars. 

\begin{figure}
\centering
\includegraphics[width=3.5in]{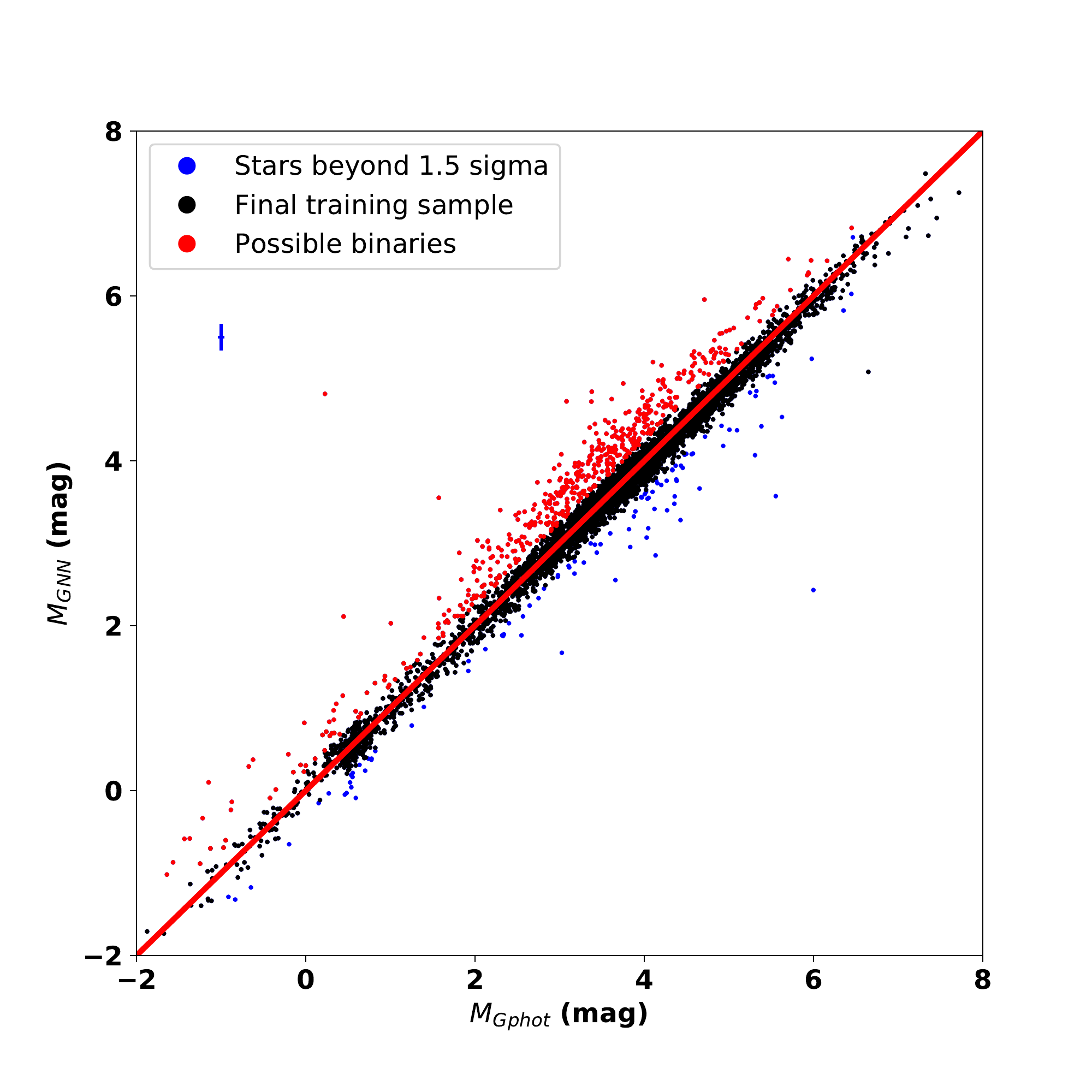}
\caption{The comparison of  photometric absolute magnitudes ($M_{Gphot}$, estimated based on distance modulus) and absolute magnitudes derived from LAMOST spectra ($M_{GNN}$) for Gaia\,EDR3 $G$ band of the initial 6,000 training stars. Black dots are the stars in the final  LGMWAS training sample. Red dots are the possible binary stars. Blue dots are stars beyond 1.5 sigma. Typical errors of $M_{Gphot}$ and $M_{GNN}$ are indicated by the error bars (blue cross) at the top left.} 
\label{train_gaia}
\end{figure}

Using the aforementioned neural network models based on the LGMWAS training stars, we estimate the aforementioned absolute magnitudes for training stars.  Fig.\,\ref{train_gaia} shows the comparison of  photometric absolute magnitudes (estimated based on distance modulus) and absolute magnitudes derived from LAMOST spectra for Gaia\,EDR3 $G$ band of our LGMWAS training stars.  There are some stars (red dots) of which the photometric absolute magnitudes are brighter than that derived from LAMOST spectra. These stars are most likely  binary stars.  In order to obtain a more reliable relation between LAMOST spectra and absolute magnitudes, we exclude these possible binary stars from our LGMWAS training sample. We also exclude stars (blue  dots) beyond 1.5 sigma   through sigma clipping process except stars with $M_{G} > 6.5$.  Then new final LGMWAS training stars (black dots) are adopted to build up the final neural network models.  

We adopt the final LGMWAS training sample to obtain the neural network models between the LAMOST spectra and aforementioned absolute magnitudes, and then estimate  absolute magnitudes using LAMOST spectra for test stars. We find that the photometric absolute magnitudes (estimated based on distance modulus) and absolute magnitudes derived from LAMOST spectra could match with each other very well.
The standard deviations of the residuals are only 0.132, 0.127, 0.136, 0.139, 0.139, 0.134, 0.157, 0.143, 0.148, 0.232, 0.143, 0.153, 0.138 and 0.143\,mag for the   $G, Bp, Rp$ of Gaia, $J, H, Ks$ of 2MASS,  $B, V, r$ of APASS, $g, r, i$ of SDSS and $W1, W2$ of WISE bands, respectively.  The systematic differences between photometric absolute magnitudes   and that  derived from LAMOST spectra are very small, almost all of them are smaller than 0.02\,mag.  One can see Fig.\,\ref{test_phot} in Appendix for more details of the comparisons. 

It is noted that the three training sets have few common sources.  The LAMOST-PASTEL and LAMOST-APOGEE have 117 common stars, LAMOST-APOGEE and  LGMWAS have 116 common stars and LAMOST-PASTEL and  LGMWAS only have one common star. Because of that one parameter has one individual neural network model,  which suggest that the stellar parameters are estimated independently, the few common stars do not affect the quality of the estimated stellar parameters. 

The catalogs of the three training sets in FITS format will be published in   \url{http://www.lamost.org/dr8/v1.0/doc/vac} together with our final value-added catalog.

\section{Applying  neural-network models to  LAMOST DR8 low-resolution spectra }
After building up neural network models based on aforementioned three training sets, we apply them to all LAMOST DR8  low-resolution spectra and estimate the absolute magnitudes, effective temperatures, surface gravity, metallicity and  chemical element abundances.  After estimating these stellar parameters, we study the quality of our results. 

In this section, we will first estimate  the uncertainties of the estimated stellar parameters using neural network method through internal comparison.  Secondly, we further check the quality of our estimates through external comparisons for all kinds of stellar parameters. The external comparisons will also help us to check whether our estimations exist systematic errors.  

\subsection{The uncertainties of the stellar parameters estimated using neural network method}
In this section, we examine the uncertainties of the  stellar parameters estimated using neural network method.  The uncertainties of these deduced stellar parameters are dependent on the spectral noise (random errors) and method errors.  It is noted that the uncertainties are also  dependent on stellar parameters as discussed  in the following chapters. We do not discuss the dependence  on stellar parameters of the uncertainties here.

\begin{figure*}
\centering
\includegraphics[width=5.5in]{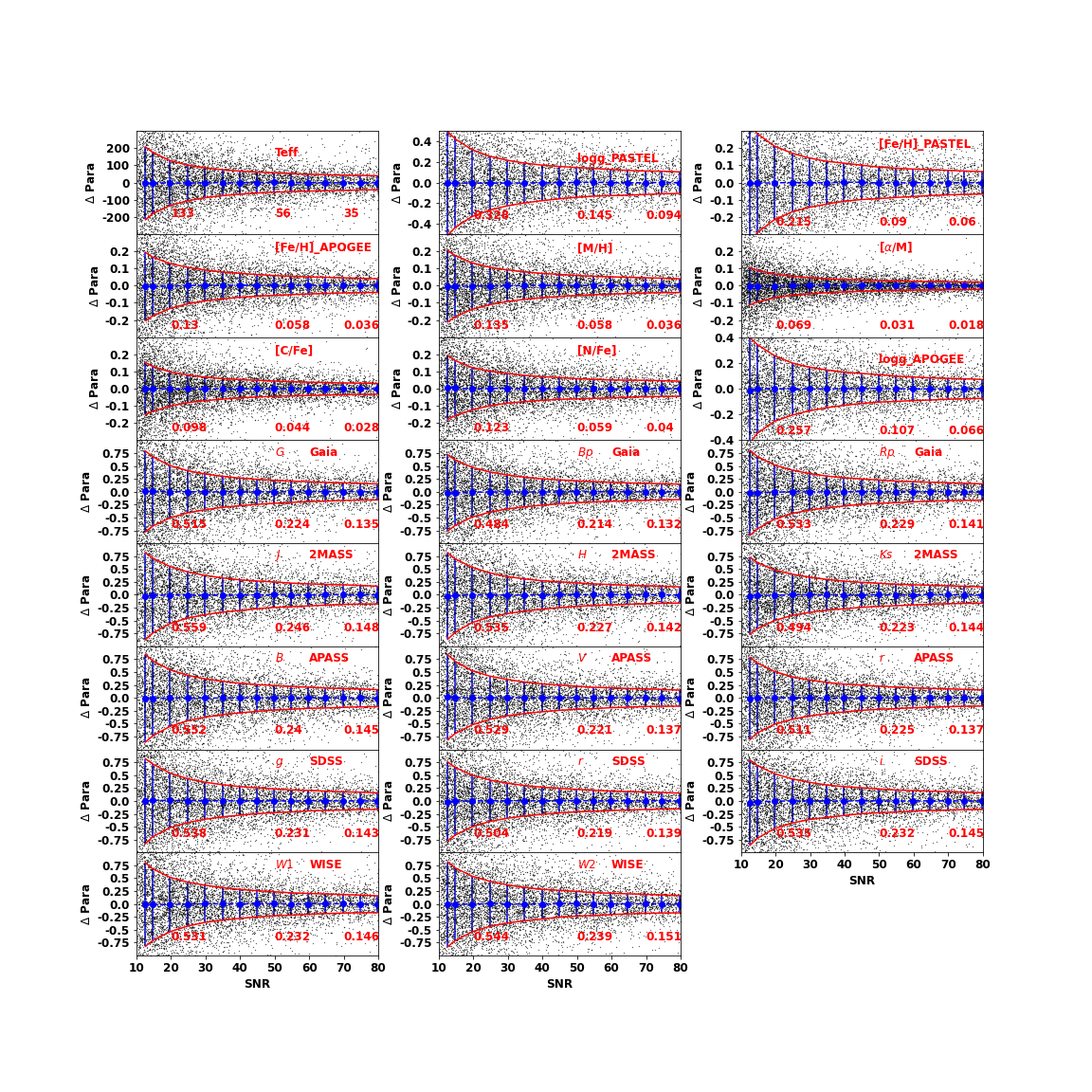}
\caption{Relative internal residuals of our stellar parameters  given by duplicate observations of similar spectral SNRs at different spectral SNR bins.  Black dots are the differences of  duplicate observations of SNRs
differences smaller than 10\,\%. Blue dots and error bars represent the medians and standard deviations (after divided by  $\sqrt{2}$) of the relative residuals in the individual spectral SNR bins. Red lines indicate fits of the standard deviations as a function of spectral SNRs. The standard deviation values of relative internal residuals at SNR $=20, 50, 100$ are respectively shown in the bottom of each panel from left to right.}
\label{para_snr}
\end{figure*}

The random errors of derived stellar parameters are estimated through comparing  results derived from duplicate observations of similar spectral SNRs (differed by less than 10\%) collected during different nights. Fig.\,\ref{para_snr} shows the relative stellar parameter estimate residuals (after divided by $\sqrt{2}$) variations with mean spectral SNRs.  In order to obtain proper random errors of stellar parameters, we fit the relative residuals with the similar equation of \cite{Huang2020}:
\begin{equation}
\sigma_{r}=a+\frac{c}{(SNR)^{b}},
\end{equation}
where $\sigma_{r}$ represents the random error.  Note that, the relative residuals are divided by 100 when we do this fit for $T_{\rm eff}$.    Besides the random errors, method errors ($\sigma_{m}$) are
also considered when we estimate the uncertainties of stellar parameters.  The final errors are given by $\sqrt{\sigma_{r}^{2}+\sigma_{m}^{2}}$.  The method errors are provided by  the relative residuals between our estimated results and the results respectively provided by PASTEL, APOGEE or distance modulus of the responding training sample. 
The resulting fit coefficients of random errors and method errors for different stellar parameters  are presented in Table\,\ref{table_snrfit}.


\begin{deluxetable}{ccccc}
\tablecaption{The method errors and the fit coefficients of random errors  of the stellar parameters' uncertainty estimates.\label{table_snrfit}}
\tablehead{
\colhead{Parameters} & \colhead{a} & \colhead{b} &\colhead{c} &\colhead{$\sigma_{m}$}
}
\startdata
$T_{\rm eff}$&0.210&1.229&42.827&75 \\
$\log\,g$\_PASTEL&0.047&1.121&7.818&0.089 \\
$\rm [Fe/H]$\_PASTEL&0.028&1.156&5.817&0.049 \\
$\rm [Fe/H]$\_APOGEE&0.0097&0.993&2.332&0.033 \\
$\rm [M/H]$&0.011&1.029&2.677&0.032 \\
$[\alpha \rm /M]$&0.0075&1.084&1.557&0.018 \\
$\rm [C/Fe]$&0.010&1.029&1.913&0.043 \\
$\rm [N/Fe]$&0.022&1.095&2.634&0.070 \\
$\log g$\_APOGEE&0.028&1.141&6.829&0.067 \\
G\_Gaia&0.049&1.048&10.442&0.092 \\
Bp\_Gaia&0.051&1.055&9.980&0.075 \\
Rp\_Gaia&0.057&1.097&12.347&0.090 \\
J\_2MASS&0.052&1.019&10.435&0.081 \\
H\_2MASS&0.052&1.090&12.451&0.082 \\
Ks\_2MASS&0.053&1.014&9.026&0.088 \\
B\_APASS&0.053&1.062&11.734&0.096 \\
V\_APASS&0.055&1.102&12.518&0.087 \\
r\_APASS&0.053&1.063&10.905&0.095 \\
g\_SDSS&0.056&1.076&11.733&0.157 \\
r\_SDSS&0.050&1.052&10.298&0.103 \\
i\_SDSS&0.054&1.068&11.538&0.098 \\
W1\_WISE&0.052&1.037&10.468&0.097 \\
W2\_WISE&0.050&1.021&10.236&0.100\\
\enddata
\end{deluxetable}

 \subsection{Comparing our effective temperature with that provided by $Gaia$-ESO and APOGEE DR16}
 The $Gaia$-ESO Public Spectroscopic Survey is designed to obtain  high quality and high resolution spectra of some 100\,000 Milky Way stars \citep{Gilmore2012, Smiljanic2014, Worley2020}. $Gaia$-ESO observed stars using the medium-resolution spectrograph GIRAFFE ($R\,\sim$\,20\,000) and the high-resolution spectrograph UVES ($R\,\sim$47 000) mounted on the VLT.  $Gaia$-ESO DR3 have provided accurate stellar atmospheric parameters and chemical  abundances  for stars observed  by $Gaia$-ESO Survey before July 2014.  As discussed in Section 3.2, APOGEE DR16 provides accurate estimates of effective temperatures and chemical element abundances using high resolution spectra.   Thus we could test the accuracy of our effective temperature measurements by comparing our results with independent measurements from the  $Gaia$-ESO and APOGEE surveys.
 
Through cross matching  LAMOST DR8 and $Gaia$-ESO DR3, we obtain 167 common stars, of which the LAMOST spectral SNR is larger than 20, the typical uncertainties of $T_{\mathrm{eff}}$ and [Fe/H] provided by $Gaia$-ESO DR3 for FGK type stars are 55\,K\, and 0.07\,dex, respectively.   We also obtain 53,295 common stars, with LAMOST spectral SNR larger than 20, APOGEE spectral SNR  larger than 70, uncertainties of [Fe/H], $T_{\mathrm{eff}}$ and $\log g$ provided by APOGEE DR16 smaller than 0.1\,dex, 150\,K and 0.15\,dex,  and $T_{\mathrm{eff}}$ of APOGEE  higher than 4000\,K,  through cross matching APOGEE DR16 and LAMOST\,DR8.  
  
\begin{figure*}
\centering
\includegraphics[width=5.5in]{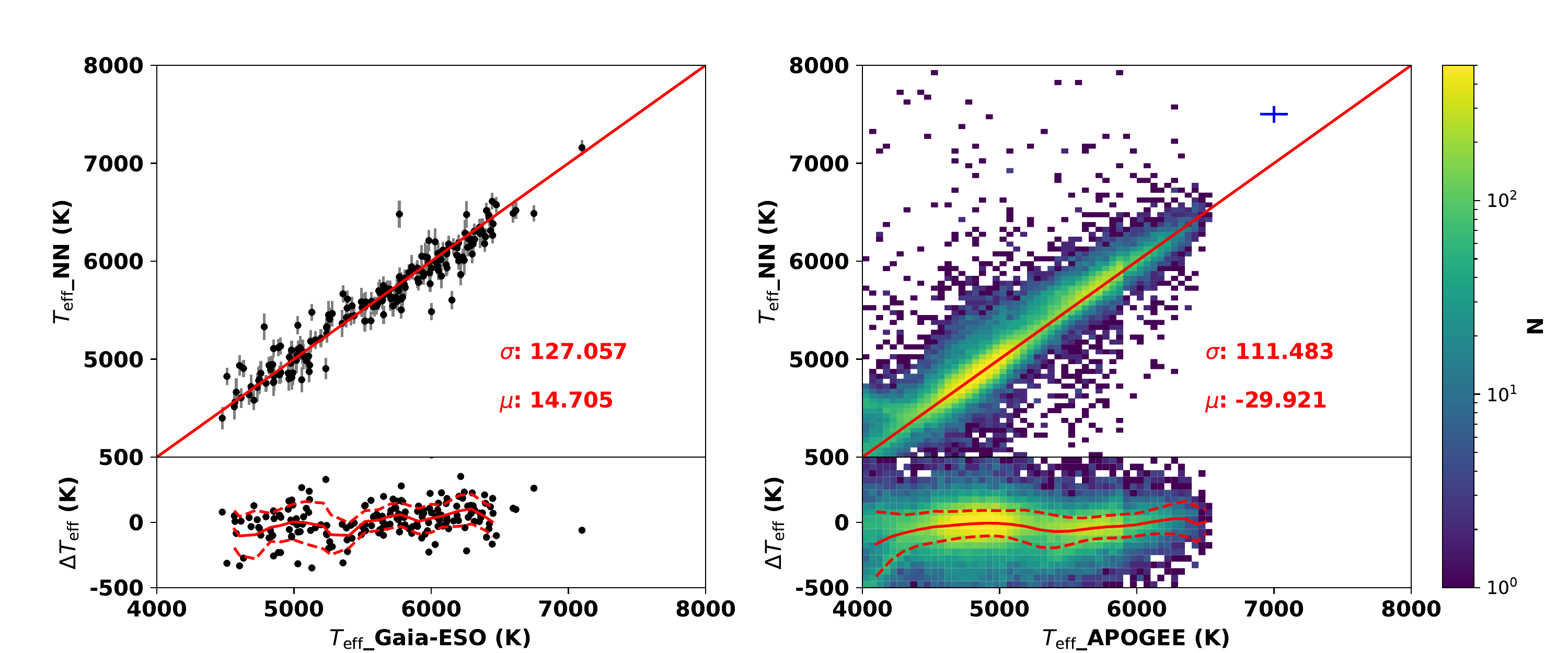}
\caption{Comparing  the effective temperatures provided by $Gaia$-ESO DR3 (left panel) and APOGEE\,DR16 (right panel) with that   derived from LAMOST spectra  based on neural-network models using LAMOST-PASTEL stars as training set.  The offset and standard deviations of the difference between the two comparisons are marked in each panel. The difference are also overplotted at the bottom of each panel.   The effective temperatures  of $Gaia$-ESO DR3 and APOGEE\,DR16 minus  that  derived from LAMOST spectra  are $\Delta T_{\mathrm{eff}}$.  Typical errors of APOGEE $T_{\mathrm{eff}}$  and our estimated $T_{\mathrm{eff}}$ are indicated by the  error bars (blue cross) in the left panel. }
\label{compare_teff}
\end{figure*}

In the Fig.\,\ref{compare_teff}, we compare  effective temperatures provided by $Gaia$-ESO DR3  and APOGEE DR16 with effective temperatures derived from LAMOST spectra based on neural-network models using LAMOST-PASTEL as training set of the current works. 
We find that the differences of  our $T_{\mathrm{eff}}$ measurements with the $Gaia$-ESO DR3  and APOGEE DR16 measurements have sigma of 127.1\,K and 111.5\,K, respectively.   The systematic differences of  our $T_{\mathrm{eff}}$ measurements with the $Gaia$-ESO DR3  and APOGEE DR16 measurements are respectively 14.7\,K and  $-29.9$\,K, which are negligible considering the typical uncertainties of stellar atmospheric parameters provided by $Gaia$-ESO DR3 and APOGEE DR16.

\subsection{Comparisons of our  surface gravity with asteroseismic measurements and that of APOGEE DR16}
In this section, we examine the quality of surface gravity measurements using the LAMOST-PASTEL ($\log\,g$-NN-PASTEL) and the LAMOST-APOGEE ($\log\,g$-NN-APOGEE) stellar sample as training sets.

In order to test the quality of estimated surface gravity, we compare our values of surface gravity with asteroseismic measurements and that of APOGEE DR16.  APOGEE DR16 provide accurate stellar parameters including accurate surface gravity.  The asteroseismic surface gravity ($\log g$) estimates inferred from the $Kepler$ data can be accurate to 0.03 dex \citep{Hekker2013, Huber2014}, which is even   much better than that  estimated with high-resolution spectroscopy ($\sim$0.1\,dex).  Cross matching $Kepler$ targets and LAMOST DR8 targets, we find $\sim$4\,000 common stars with LAMOST spectral SNR\,$> 20$ and error of asteroseismic  surface gravity Err$_{\log g} < 0.05$\,dex. 

\begin{figure*}
\centering
\includegraphics[width=5.5in]{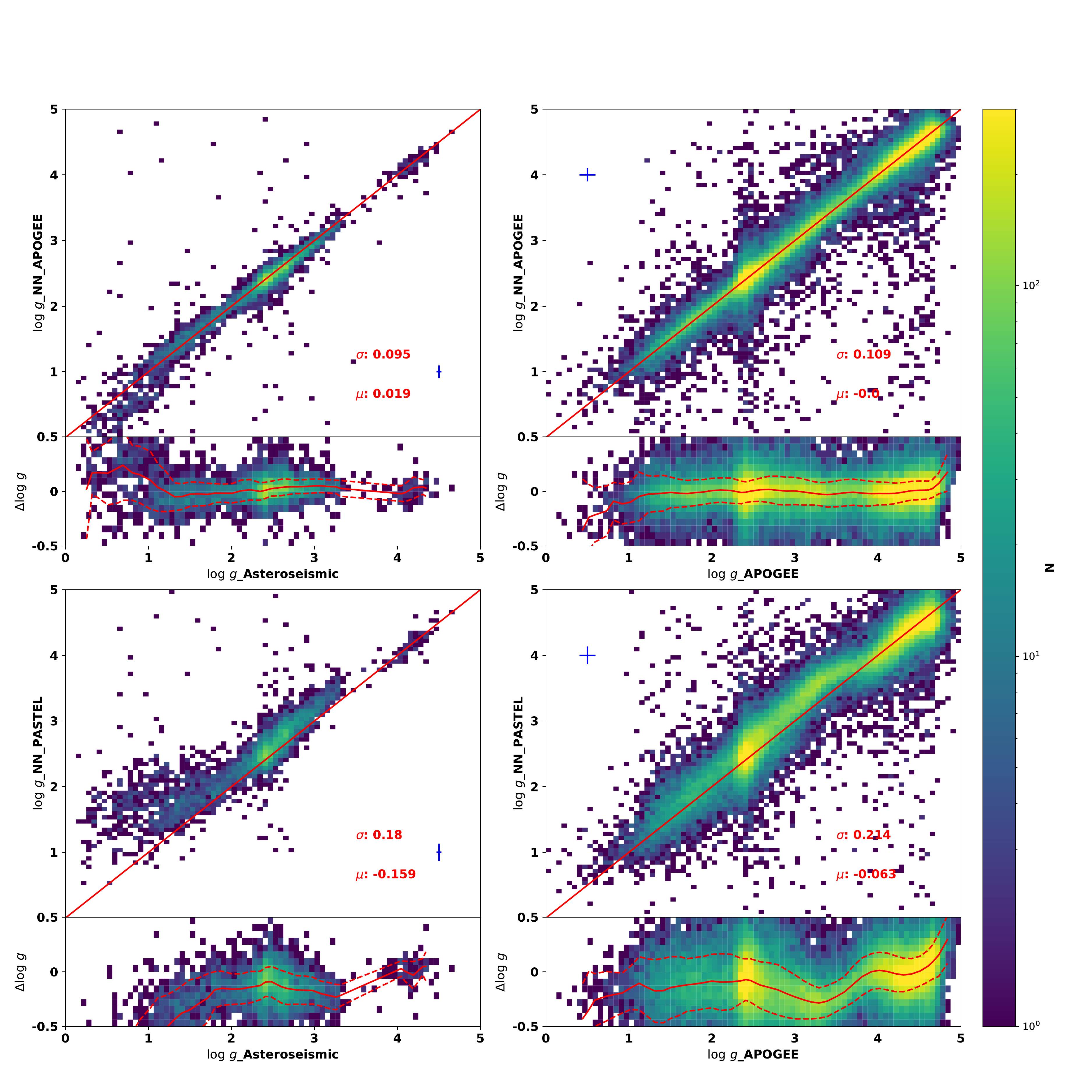}
\caption{Comparing our $\log\,g$ values estimated using LAMOST-PASTEL (bottom panels) and LAMOST-APOGEE (top panels) as training sets with  asteroseismic  $\log\,g$ (left panels)  and  $\log\,g$ from APOGEE\,DR16 (right panels).  The offset and standard deviations  are marked in  each panel. The difference are also overplotted at the bottom of each panel.   The asteroseismic  $\log\,g$ and  APOGEE $\log\,g$ minus  that  derived from LAMOST spectra  are $\Delta \log\,g$. Typical errors of asteroseismic/APOGEE $\log\,g$  and our estimated $\log\,g$ are indicated by the  error bars (blue cross) in each panel. } 
\label{compare_logg}
\end{figure*}

Fig.\,\ref{compare_logg} shows the comparisons between our two kinds of $\log\,g$ estimates and asteroseismic  $\log\,g$ and  APOGEE $\log\,g$ for common stars with LAMOST spectral SNR\,$> 20$.  
From this figure, we can find that our $\log\,g$ estimated using LAMOST-APOGEE as training set could match very well with the asteroseismic  $\log\,g$ and  APOGEE $\log\,g$.  The offset of the difference of our $\log\,g$-NN-APOGEE  with  asteroseismic  $\log\,g$ and APOGEE $\log\,g$ are negligible. The standard deviations are only 0.095 and 0.109\,dex for  the difference of our $\log\,g$-NN-APOGEE  with  asteroseismic  $\log\,g$ and APOGEE $\log\,g$, respectively. 
The $\log\,g$ estimated using LAMOST-PASTEL as training set could match very well with asteroseismic  $\log\,g$ or  APOGEE $\log\,g$ for dwarf stars. However, our $\log\,g$-NN-PASTEL for giant stars are over estimated with $\sim$ 0.25\,dex. 
The offset of the difference of our $\log\,g$-NN-PASTEL with asteroseismic $\log\,g$ and  APOGEE $\log\,g$ are $-0.159$ and $-0.063$\,dex. The standard deviations are only 0.18 and 0.214\,dex for  the difference of our $\log\,g$-NN-PASTEL with  asteroseismic our $\log\,g$ and  APOGEE $\log\,g$, respectively.  The $\log\,g$-NN-APOGEE are better than the $\log\,g$-NN-PASTEL for stars with $T_{\rm eff} < 6500$\,K, especially for giant stars.

\subsection{The external comparisons of [Fe/H] values}
In this section, we examine the quality of [Fe/H] measurements using the LAMOST-PASTEL ([Fe/H]-NN-PASTEL) and LAMOST-APOGEE ([Fe/H]-NN-APOGEE) stellar samples as training sets.  We compare our [Fe/H] values with that of $Gaia$-ESO Survey and APOGEE DR16 to verify whether our [Fe/H] have  systematic errors.  In addition, the [Fe/H] dispersions  of member stars in open  clusters  are also studied to check the uncertainties of metallicity. 

\subsubsection{Comparisons with $Gaia$-ESO Survey}
\begin{figure*}
\centering
\includegraphics[width=5.5in]{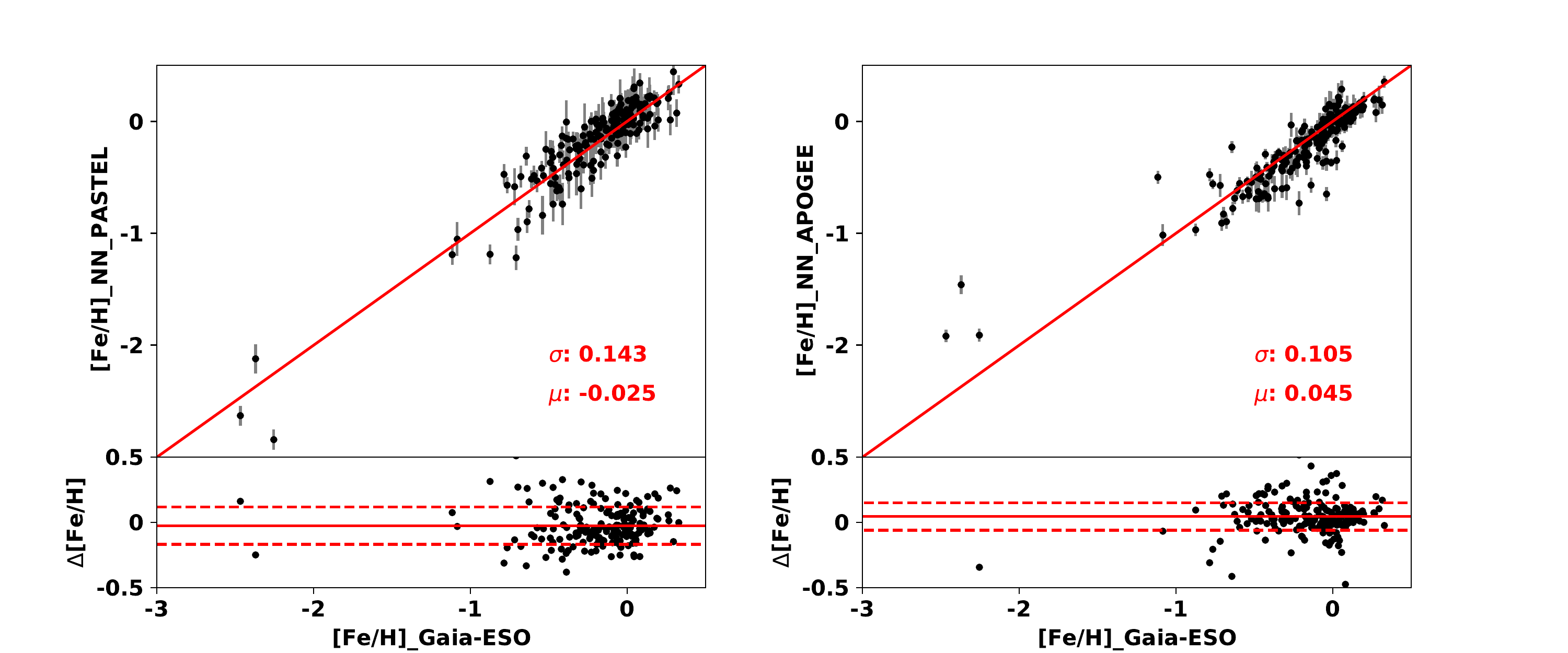}
\caption{Comparing  the [Fe/H] provided by $Gaia$-ESO DR3 with  our [Fe/H] measurements using  the LAMOST-PASTEL (left panel) and LAMOST-APOGEE (right panel) stellar sample as training sets.  The offset and standard deviations of the difference between the two comparisons are marked in each panel. The difference are also overplotted at the bottom of each panel.   The [Fe/H] values derived from LAMOST spectra minus that  provided by $Gaia$-ESO DR3  are $\Delta \rm [Fe/H]$. The horizontal lines and dashed lines indicate $\mu$ and $\mu \pm \sigma$, respectively. }
\label{compare_feh_gaiaeso}
\end{figure*}

 As discussed in section 4.2, the $Gaia$-ESO DR3 have derived  accurate stellar atmospheric parameters and chemical element  abundances using the medium and high-resolution spectra.  We test the accuracy of the two kinds of [Fe/H] measurements through comparing with that provided by  $Gaia$-ESO DR3.   Fig.\,\ref{compare_feh_gaiaeso} shows the comparisons between our two kinds of [Fe/H] measurements with [Fe/H] provided by  $Gaia$-ESO DR3.   We find that the differences of [Fe/H] values provided by $Gaia$-ESO DR3  with [Fe/H]-NN-PASTEL and  [Fe/H]-NN-APOGEE  have standard deviation of 0.143\,dex and 0.105\,dex, respectively.   The systematic differences of  [Fe/H] of  $Gaia$-ESO DR3 with [Fe/H]-NN-PASTEL and  [Fe/H]-NN-APOGEE are respectively $-0.025$\,dex and $0.045$\,dex. 
 The comparison suggests that our  [Fe/H] measurements using the LAMOST-PASTEL and LAMOST-APOGEE stellar samples as training sets  could match well with the [Fe/H] values of $Gaia$-ESO.

\subsubsection{Comparison with APOGEE\,DR16}
\begin{figure*}
\centering
\includegraphics[width=5.5in]{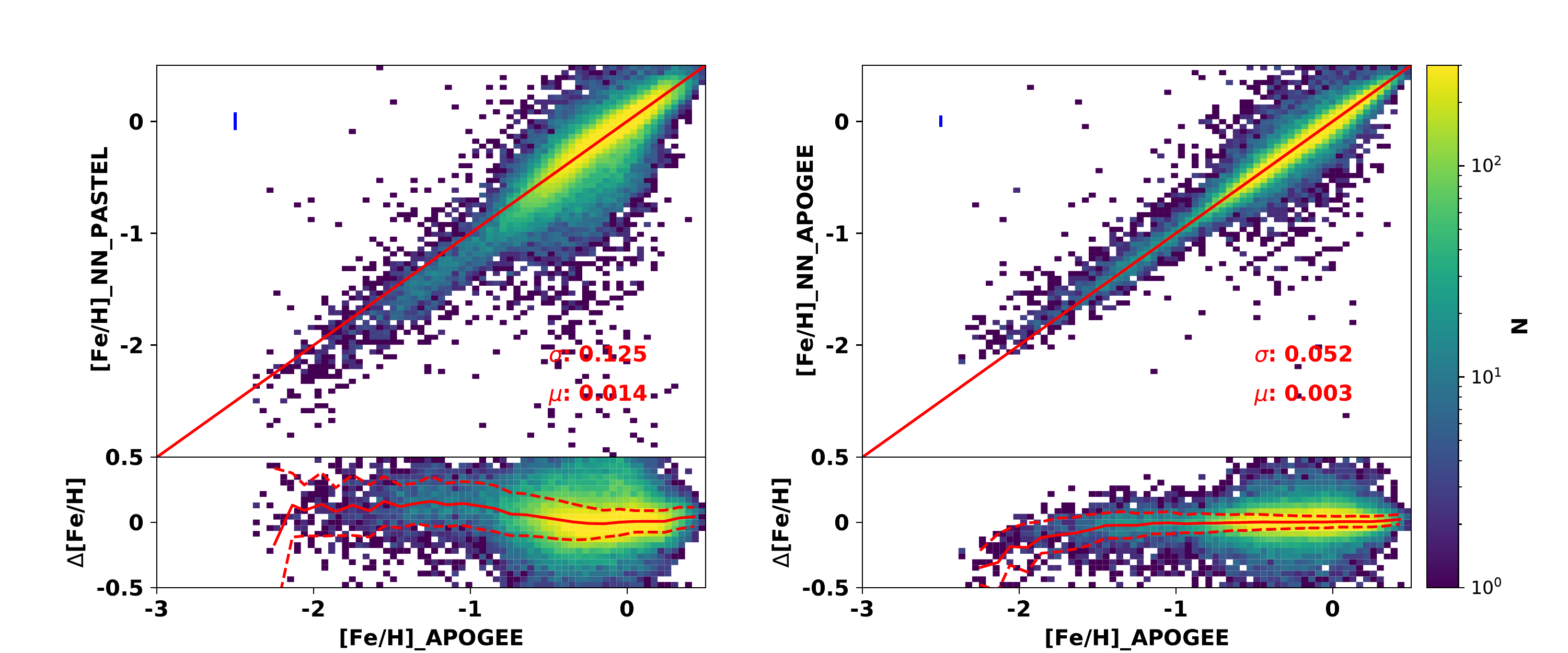}
\caption{Similar with Fig.\,\ref{compare_feh_gaiaeso}, but for the comparisons with the [Fe/H] values of  APOGEE\,DR16.  Typical errors of our two kinds of [Fe/H] and APOGEE [Fe/H] are indicated by the  error bars (blue cross) in each panel. }
\label{compare_feh_apogee}
\end{figure*}

We also compare our two kinds of [Fe/H] measurements with that provided by APOGEE\,DR16 in order to test the quality of our [Fe/H] values. Fig.\,\ref{compare_feh_apogee} shows the comparisons.   We find that the differences of [Fe/H] values provided by APOGEE  with [Fe/H]-NN-PASTEL and  [Fe/H]-NN-APOGEE  have standard deviations of 0.125\,dex and 0.052\,dex, respectively.   The systematic differences of  [Fe/H] of  APOGEE with [Fe/H]-NN-PASTEL and  [Fe/H]-NN-APOGEE are respectively $0.014$\,dex and $0.003$\,dex. 
The comparison suggests that our  [Fe/H] measurements using the LAMOST-PASTEL and LAMOST-APOGEE stellar samples as training sets  could match well with the [Fe/H] values of APOGEE\,DR16. 

\subsubsection{The [Fe/H] values of  the candidates of open cluster members}
Stars in open cluster (OC) are believed to form almost simultaneously from a single gas cloud, thus they have almost the same metallicity. Thus,  OC is a good test bed to check the uncertainty of metallicity determinations.  8,811 cluster member stars observed by LAMOST have been provided by  \cite{Zhongjing2020} and \cite{Zhongjingcatalog}, who cross-match  the cluster member stars of OCs provided by \cite{Cantat-Gaudin2018} with LAMOST\,DR5.  Here, we simply test the uncertainties of our metallicity  estimates using these cluster members.   We select cluster members in four OCs  of $\rm Melotte\,20$, $\rm Melotte\,22$, $\rm NGC\,2632$, $\rm NGC\,2682$ (there are enough number of cluster members targeted by LAMOST in these four OCs) with spectral SNR\,$> 30$ and $4,000 < T_{\mathrm{eff}} < 8,500\,$K to do the test. 

\begin{figure*}
\centering
\includegraphics[width=5.5in]{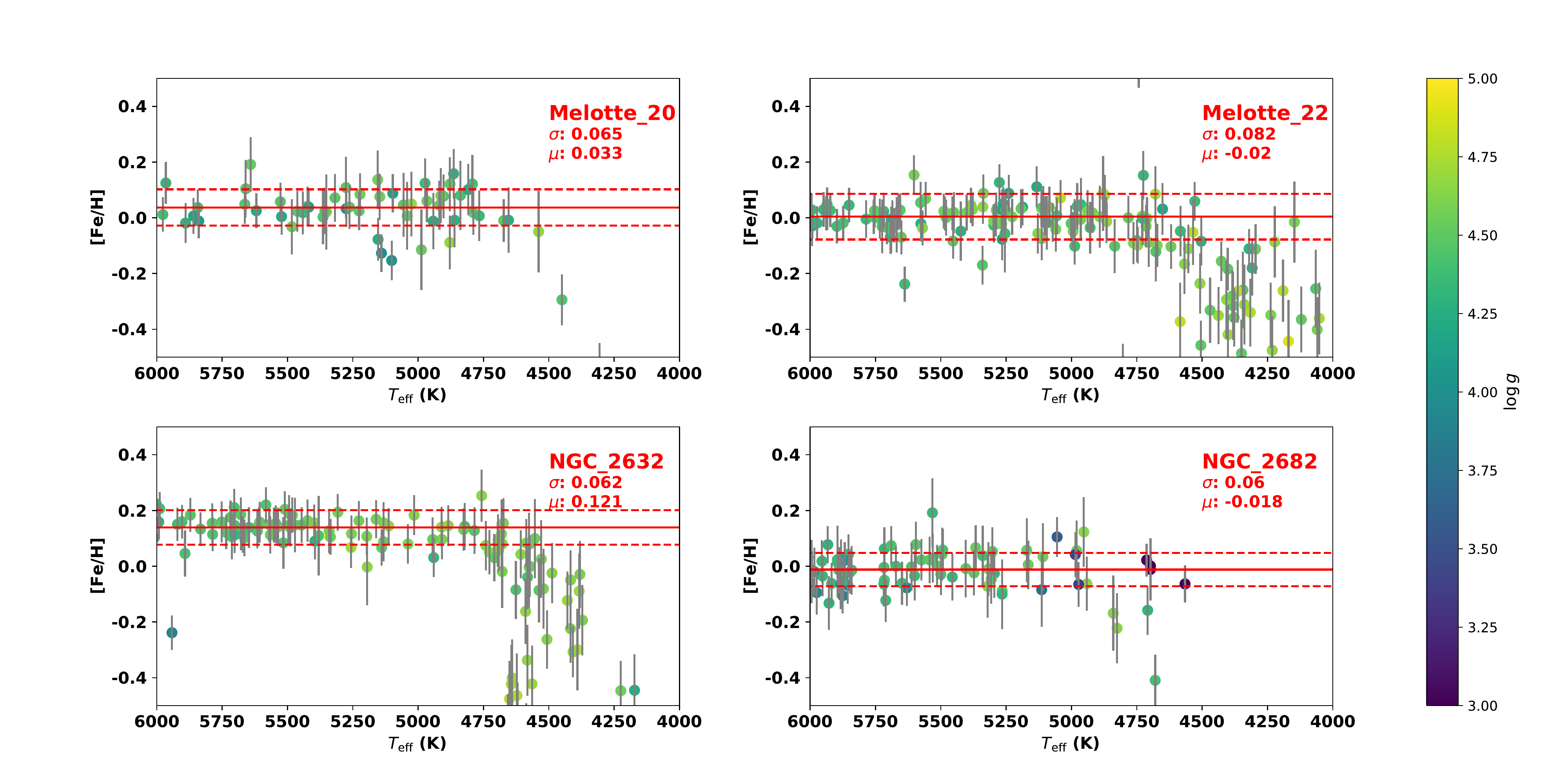}
\caption{The metallicity (estimated using LAMOST-PASTEL common stars as training sample) variations with $T_{\mathrm{eff}}$  of cluster members in the  four OCs of  $\rm Melotte\,20$, $\rm Melotte\,22$, $\rm NGC\,2632$, $\rm NGC\,2682$. The mean value $\mu$ and standard deviations $\sigma$ of the metallicity are marked in each panel. The red line indicates the mean value $\mu$, the red dashed lines indicate the $\mu$ minus/plus $\sigma$ dex. The gray lines show the error of each metallicity estimate.}
\label{feh_pastel_cluster}
\end{figure*}

\begin{figure*}
\centering
\includegraphics[width=5.5in]{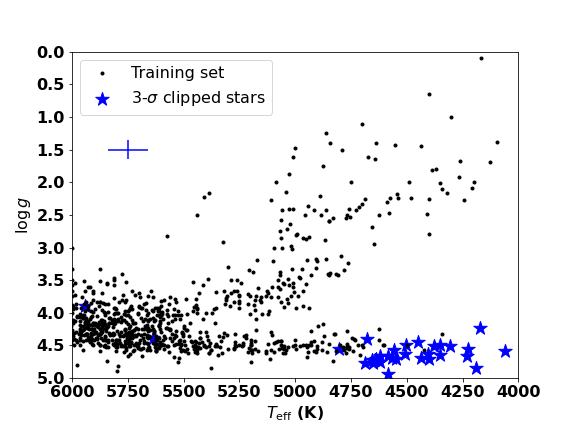}
\caption{The distributions of LAMOST-PASTEL training set (black dots) and stars in these four OCs with significantly underestimated [Fe/H] values (blue stars) in the plane of $T_{\rm eff}$--$\log\,g$. The stars with significantly underestimated [Fe/H] values are  3-$\sigma$ clipped stars. Typical errors of $T_{\rm eff}$ and $\log g$ are  indicated by the  error bars (blue cross) at the top left of the figure.}
\label{feh_pastel_outlier}
\end{figure*}

Fig.\,\ref{feh_pastel_cluster} shows the [Fe/H] (using LAMOST-PASTEL common stars as training set) variations with $T_{\mathrm{eff}}$ for cluster members  in the aforementioned four OCs.  From this figure, we can find that the standard deviations of the metallicity distributions of cluster members in these four OCs are almost similar with (or slightly larger than) the uncertainties of our metallicity determinations.  The metallicity have no trend with $T_{\mathrm{eff}}$  when $T_{\mathrm{eff}} > $4\,500--4\,750\,K. The metallicity of  stars with $T_{\mathrm{eff}} < $4\,500--4\,750\,K are underestimated, most of these stars are dwarf stars (cool dwarf stars). As shown in Fig.\,\ref{parameter_space} and Fig.\,\ref{feh_pastel_outlier}, the LAMOST-PASTEL training sample  only has few cool dwarf stars. Stars with significantly underestimated [Fe/H] values are almost located outside the LAMOST-PASTEL training parameter grid as shown in Fig.\,\ref{feh_pastel_outlier}.  

\begin{figure*}
\centering
\includegraphics[width=5.5in]{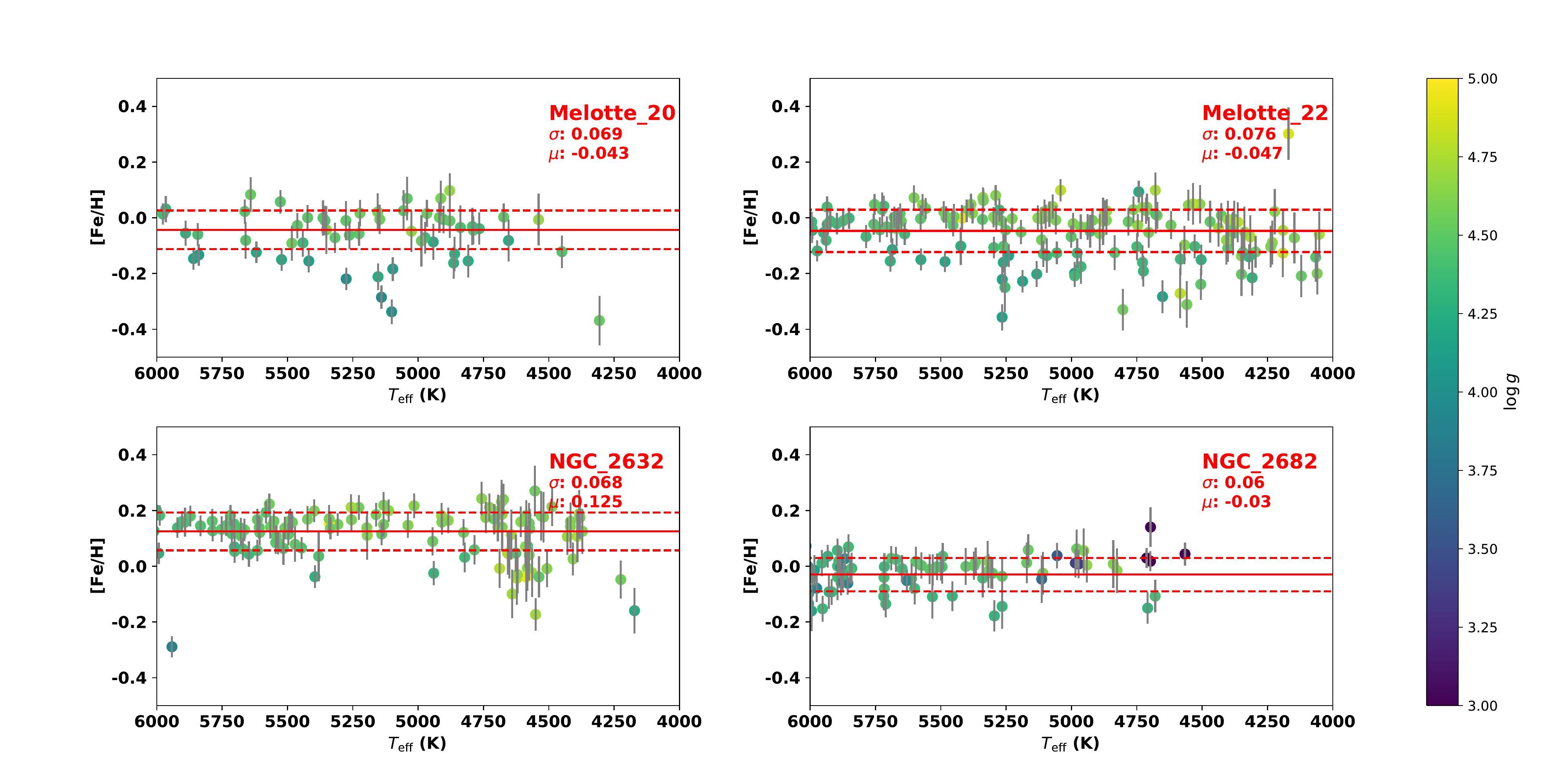}
\caption{Similar with Fig.\,\ref{feh_pastel_cluster}, but for the results of [Fe/H] estimations using LAMOST-APOGEE common stars as training set. }
\label{feh_apogee_cluster}
\end{figure*}

Fig.\,\ref{feh_apogee_cluster} shows the [Fe/H]  (using LAMOST-APOGEE common stars as training set) variations with $T_{\rm eff}$ of cluster members in  the aforementioned four  OCs. We can find that the standard deviations of the metallicity distributions of these cluster members are almost similar with the uncertainty  of our metallicity determinations.  The metallicity have no trend with $T_{\mathrm{eff}}$ when $T_{\mathrm{eff}} > $\,4\,500--4\,750\,K for cluster members in these four OCs.  
The metallicity of  stars with $T_{\mathrm{eff}} < $\,4\,500--4\,750\,K are slightly underestimated for stars in these four OCs. 
Similar with the results of [Fe/H] values estimated using LAMOST-PASTEL common stars as training set, the underestimations of metallicity may be the consequence of the lack of cool dwarf training stars.  The metallicity underestimates of cool dwarfs  here are much better than that estimated using LAMOST-PASTEL common stars as training set, through comparing Fig.\,\ref{feh_pastel_cluster} and Fig.\,\ref{feh_apogee_cluster}. 

\subsection{The external comparisons of [M/H],  [$\alpha$/M], [C/Fe] and [N/Fe] values}
In this section, we examine the quality of [M/H],  [$\alpha$/M], [C/Fe] and [N/Fe] measurements using  LAMOST-APOGEE stellar sample as training set.   We  first compare our  chemical abundance values with that of APOGEE DR16 to study the uncertainties and systematic errors of them.  The chemical abundance dispersions  of member stars in OCs  are also studied to check the uncertainties of them. 

\subsubsection{Comparisons  with APOGEE DR16}
We compare our estimates of [M/H],  [$\alpha$/M], [C/Fe] and [N/Fe] with that of APOGEE DR16 in order to study the quality of our estimates. 

\begin{figure*}
\centering
\includegraphics[width=5.5in]{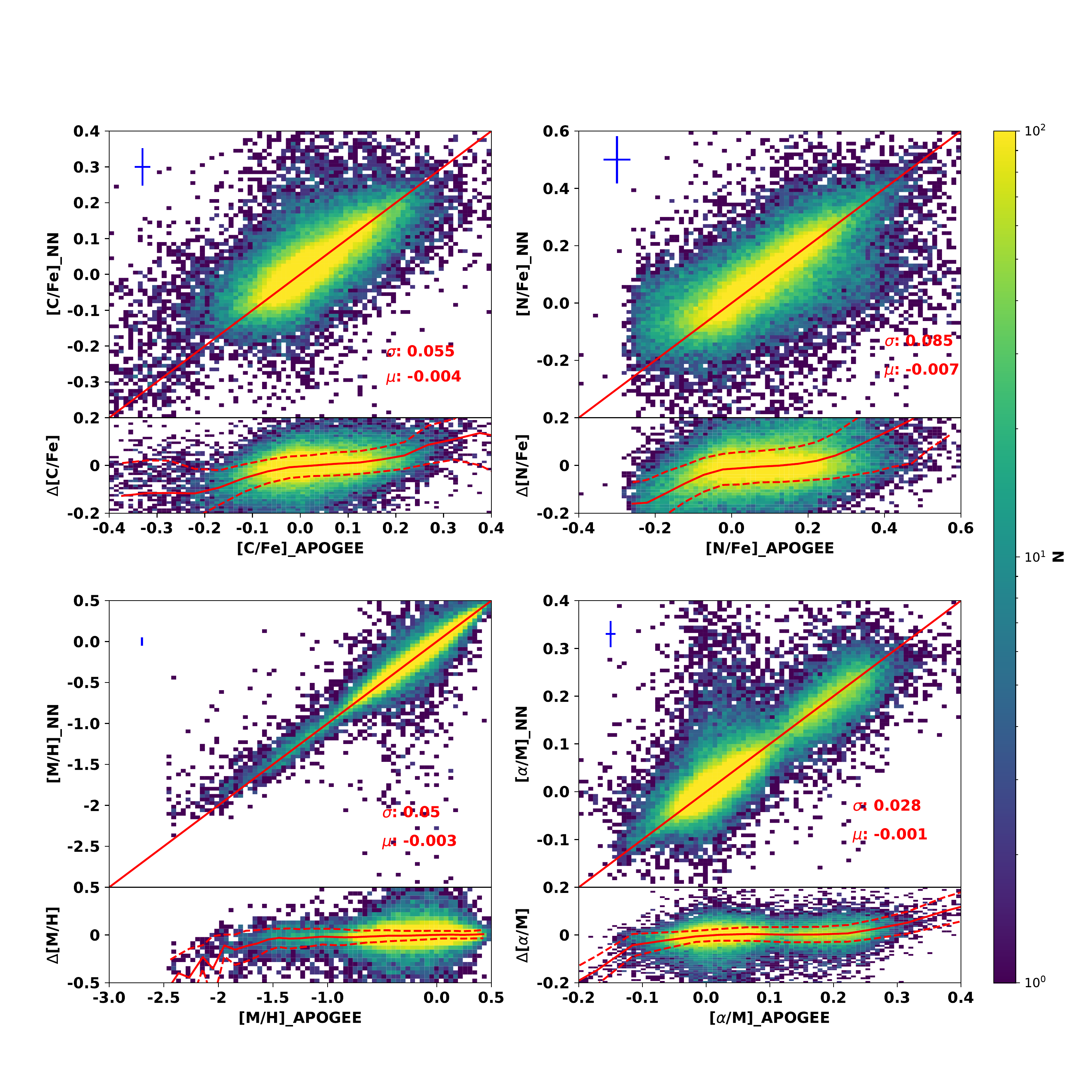}
\caption{Comparisons between the chemical abundance estimates provided by APOGEE DR16 and that  derived from LAMOST spectra  based on the neural network method using LAMOST-APOGEE common stars as training set.  The offset and standard deviations of the difference between the two kinds of  chemical abundance  measurements are marked in each panel. The difference are also overplotted at the bottom of each panel.   The chemical abundance  of APOGEE DR16 minus  that  derived from LAMOST spectra  are $\Delta$[M/H], $\Delta[\alpha/\rm M]$, $\Delta$[C/Fe] or $\Delta$[N/Fe]. Typical errors of the chemical abundance of us and APOGEE are  indicated by the  error bars (blue cross) at the top left of each panel.}
\label{compare_abun_apogee}
\end{figure*}

Fig.\,\ref{compare_abun_apogee} shows the comparison results.   
Our estimates of [M/H],  [$\alpha$/Fe], [C/Fe] and [N/Fe]  could match well with those provided by APOGEE DR16.   We find that the differences of  [M/H],  [$\alpha$/Fe], [C/Fe] and [N/Fe] between our estimates and those of APOGEE DR16 have standard deviations of 0.05\,dex, 0.028\,dex, 0.055\,dex and 0.085\,dex,  respectively.   The systematic differences of [M/H],  [$\alpha$/Fe], [C/Fe] and [N/Fe] between the two estimates are respectively $-0.003$\,dex, $-0.001$\,dex, $-0.004$\,dex and $-0.007$\,dex.  The offsets are so small that could be ignored.   These results are suitable for LAMOST spectra with spectral SNR\,$> 20$, the dispersion  of the difference will be much smaller for the results of stars with higher quality LAMOST spectra.

 \subsubsection{The chemical abundances of the candidates of open cluster members}
As discussed in Section\,4.4.3, OC  is a good test bed to check the uncertainties of chemical abundance determinations. Here, we  briefly study the uncertainties of chemical abundances  estimated by our neural network models using LAMOST-APOGEE common stars as training set of LAMOST cluster members identified by \cite{Zhongjing2020} and \cite{Zhongjingcatalog}. 

\begin{figure*}
\centering
\includegraphics[width=5.5in]{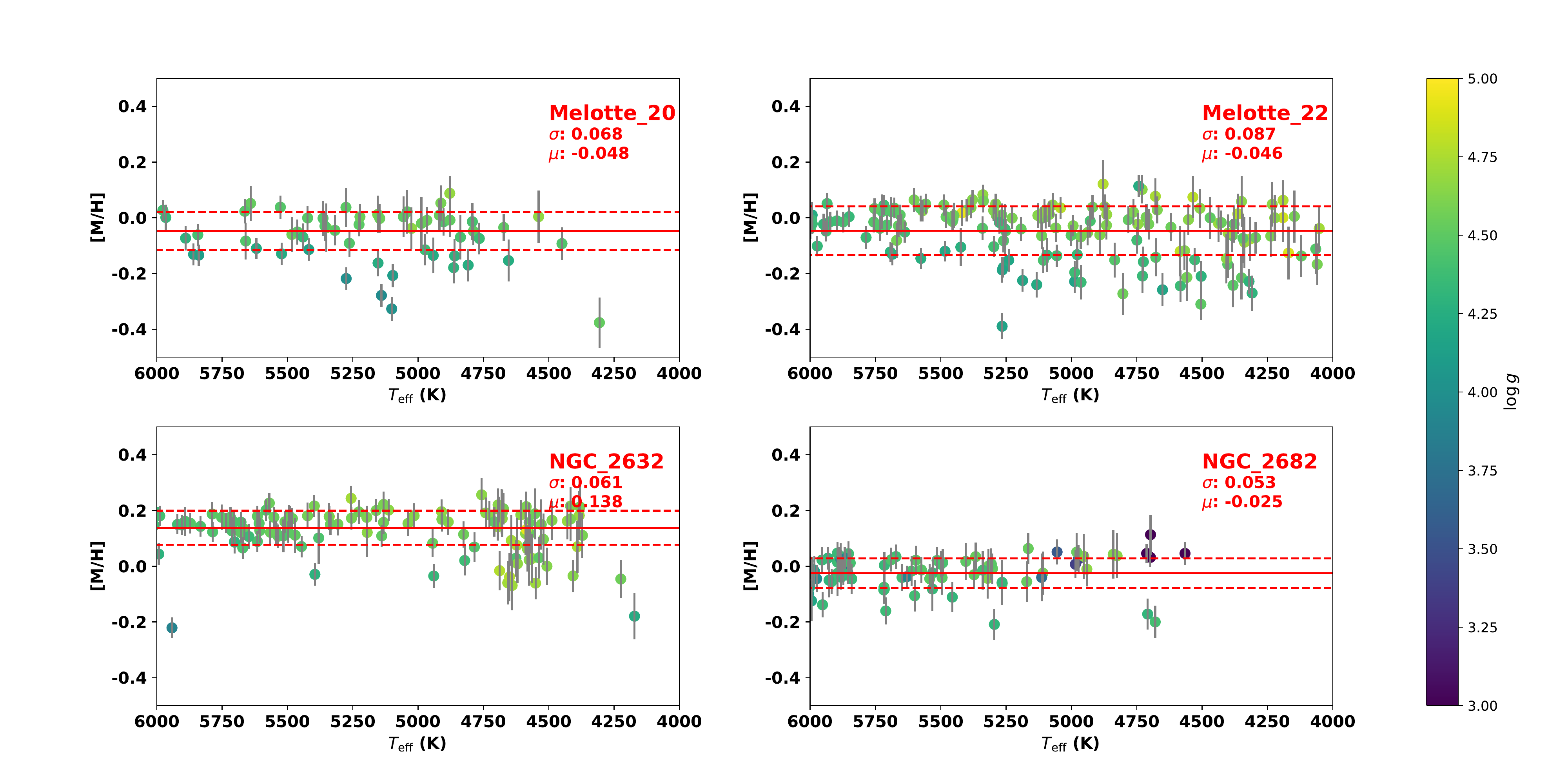}
\caption{Similar with  Fig.\,\ref{feh_pastel_cluster}, but for results of [M/H] estimations using LAMOST-APOGEE common stars as training set. }
\label{mh_apogee_cluster}
\end{figure*}

\begin{figure*}
\centering
\includegraphics[width=5.5in]{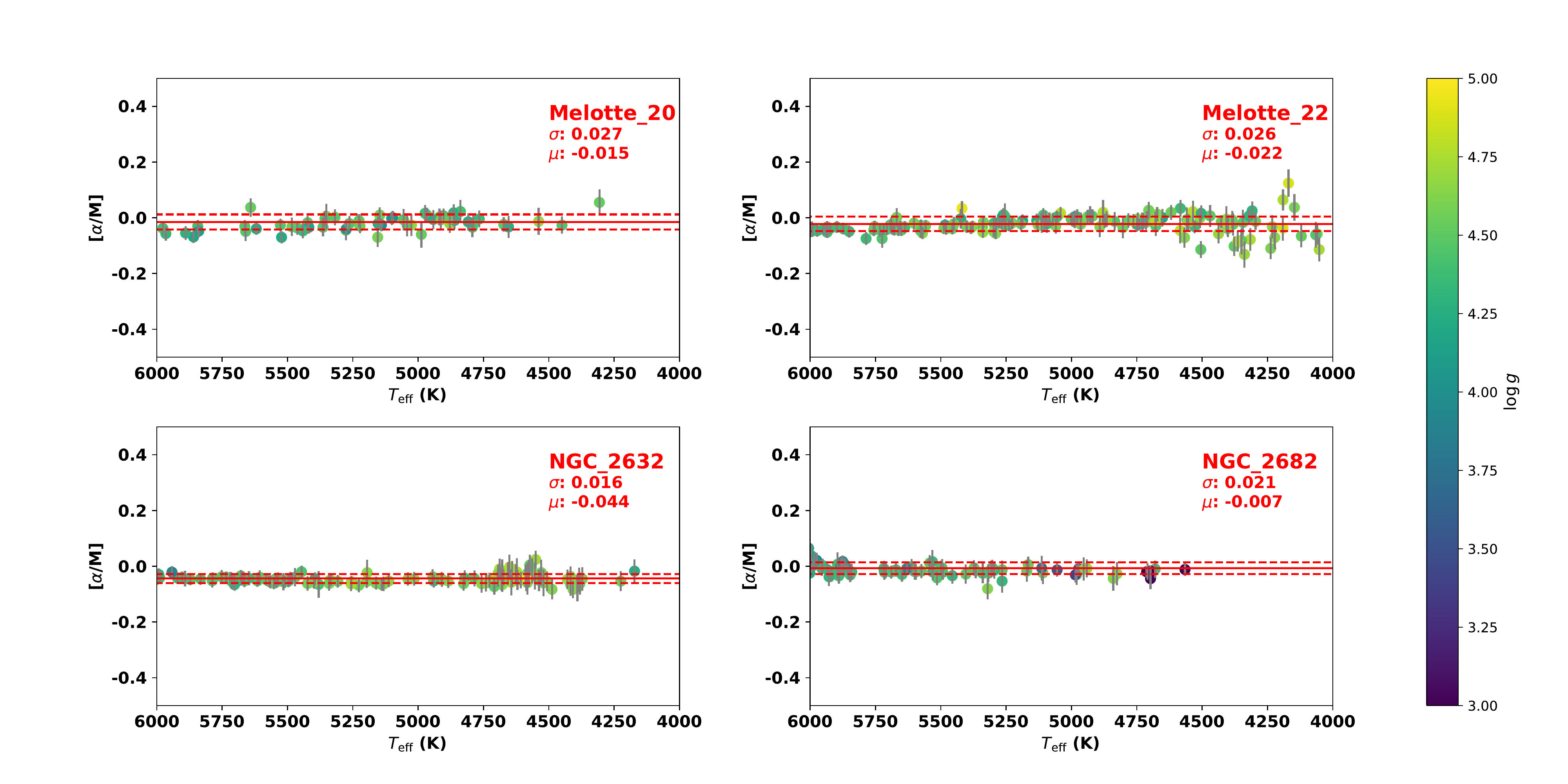}
\caption{Similar with  Fig.\,\ref{feh_pastel_cluster}, but for results of [$\alpha$/M] estimations using LAMOST-APOGEE common stars as training set.}
\label{afe_apogee_cluster}
\end{figure*}

\begin{figure*}
\centering
\includegraphics[width=5.5in]{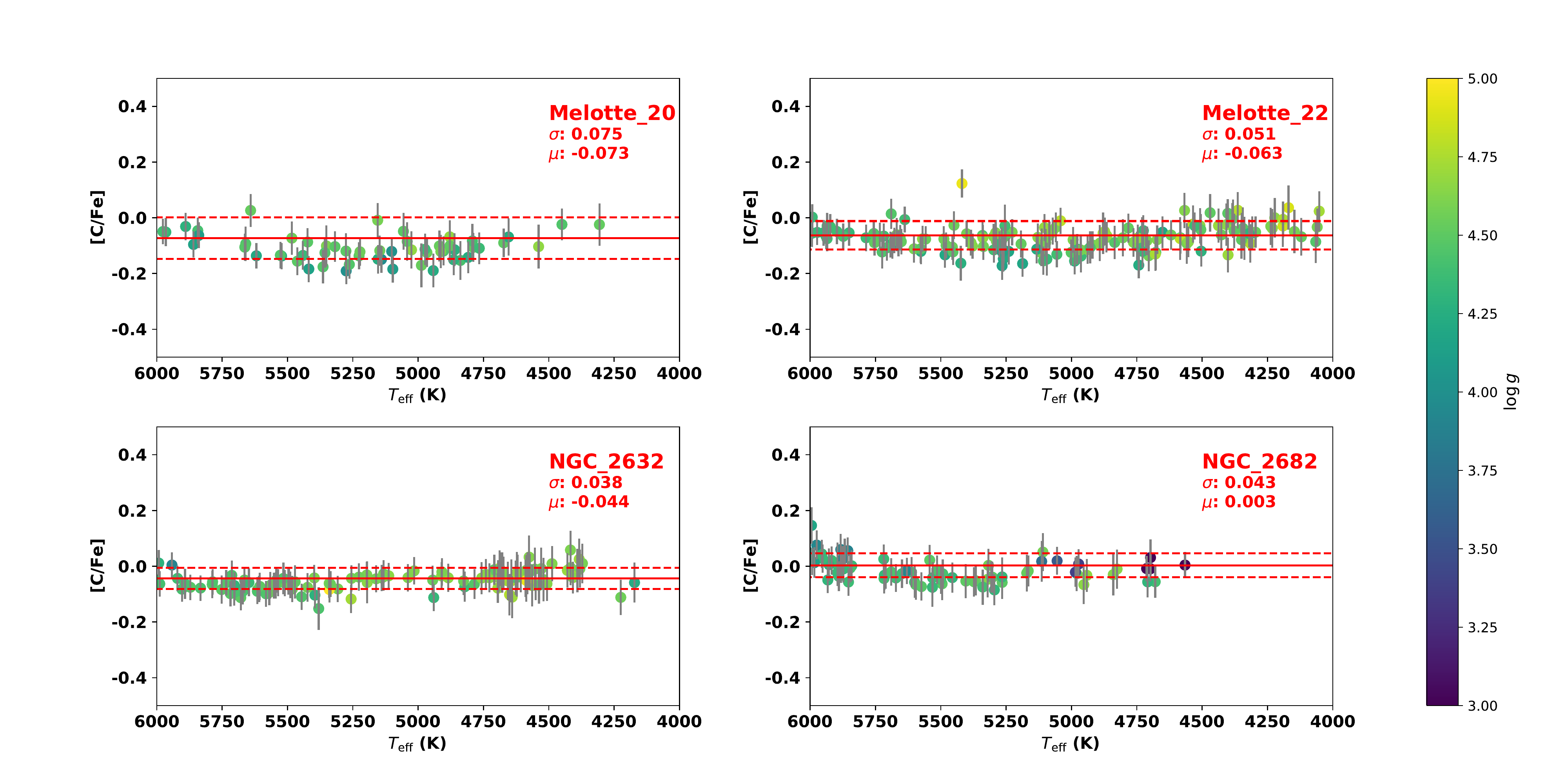}
\caption{Similar with  Fig.\,\ref{feh_pastel_cluster}, but for results of [C/Fe] estimations using LAMOST-APOGEE common stars as training set. }
\label{cfe_apogee_cluster}
\end{figure*}

\begin{figure*}
\centering
\includegraphics[width=5.5in]{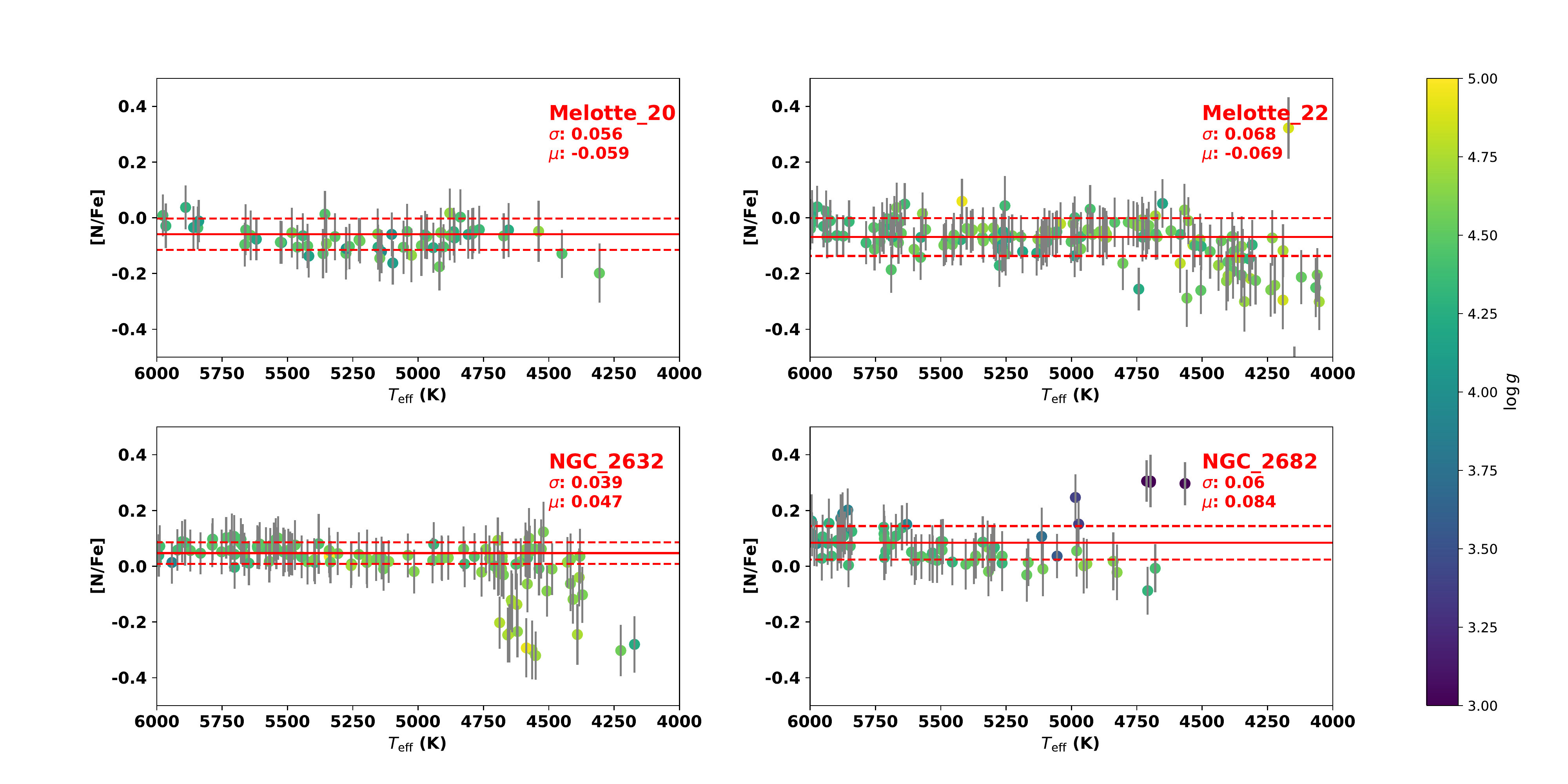}
\caption{Similar with Fig.\,\ref{feh_pastel_cluster}, but for results of [N/Fe] estimations using LAMOST-APOGEE common stars as training set. }
\label{nfe_apogee_cluster}
\end{figure*}

\begin{figure*}
\centering
\includegraphics[width=5.5in]{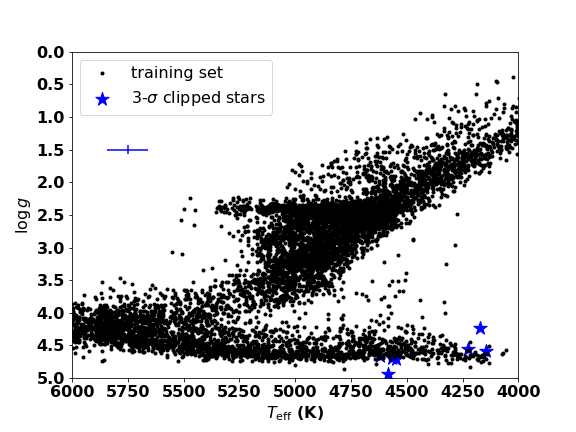}
\caption{The distributions of LAMOST-APOGEE training stars (black dots) and stars with significant underestimated [N/Fe] (blue stars)  in the $T_{\rm eff}$--$\log\,g$ plane. The stars with significantly underestimated [N/Fe] values are  3-$\sigma$ clipped stars. Typical errors of $T_{\rm eff}$ and $\log g$ are  indicated by the  error bars (blue cross) at the top left of the figure.}
\label{outlier_nfe_apogee}
\end{figure*}

Fig.\,\ref{mh_apogee_cluster}, Fig.\,\ref{afe_apogee_cluster}, Fig.\,\ref{cfe_apogee_cluster} and Fig.\,\ref{nfe_apogee_cluster} show respectively, the  [M/H], [$\alpha$/M], [C/Fe] and [N/Fe] variations with $T_{\rm eff}$ of stellar members in the aforementioned four  OCs.  The results of [M/H] are similar with our [Fe/H]  using LAMOST-APOGEE as training set.  
The variations  of [$\alpha$/M], [C/Fe] and [N/Fe]  with $T_{\rm eff}$  are very small, except for the underestimated  [N/Fe] of  stars with $T_{\rm eff} < 4,500$\,K. The underestimations of  [N/Fe]  of  stars with $T_{\rm eff} < 4,500$\,K may be the consequence of lack of cool dwarf stars in the training set as discussed in section 4.4.3.  Fig.\,\ref{outlier_nfe_apogee} shows the distributions of stars with significant underestimated [N/Fe]  in the $T_{\rm eff}$--$\log\,g$ plane. We can find that these stars are almost located  outside the training parameter grid or at the edge of the training parameter grid.
From Fig.\,\ref{mh_apogee_cluster}, Fig.\,\ref{afe_apogee_cluster}, Fig.\,\ref{cfe_apogee_cluster} and Fig.\,\ref{nfe_apogee_cluster}, one can see that the standard deviations of  [M/H], [$\alpha$/Fe], [C/Fe] and [N/Fe] of stellar members in these four OCs are very small.  Thus our estimates of [M/H], [$\alpha$/Fe], [C/Fe] and [N/Fe] are very good.   

\begin{figure*}
\centering
\includegraphics[width=5.5in]{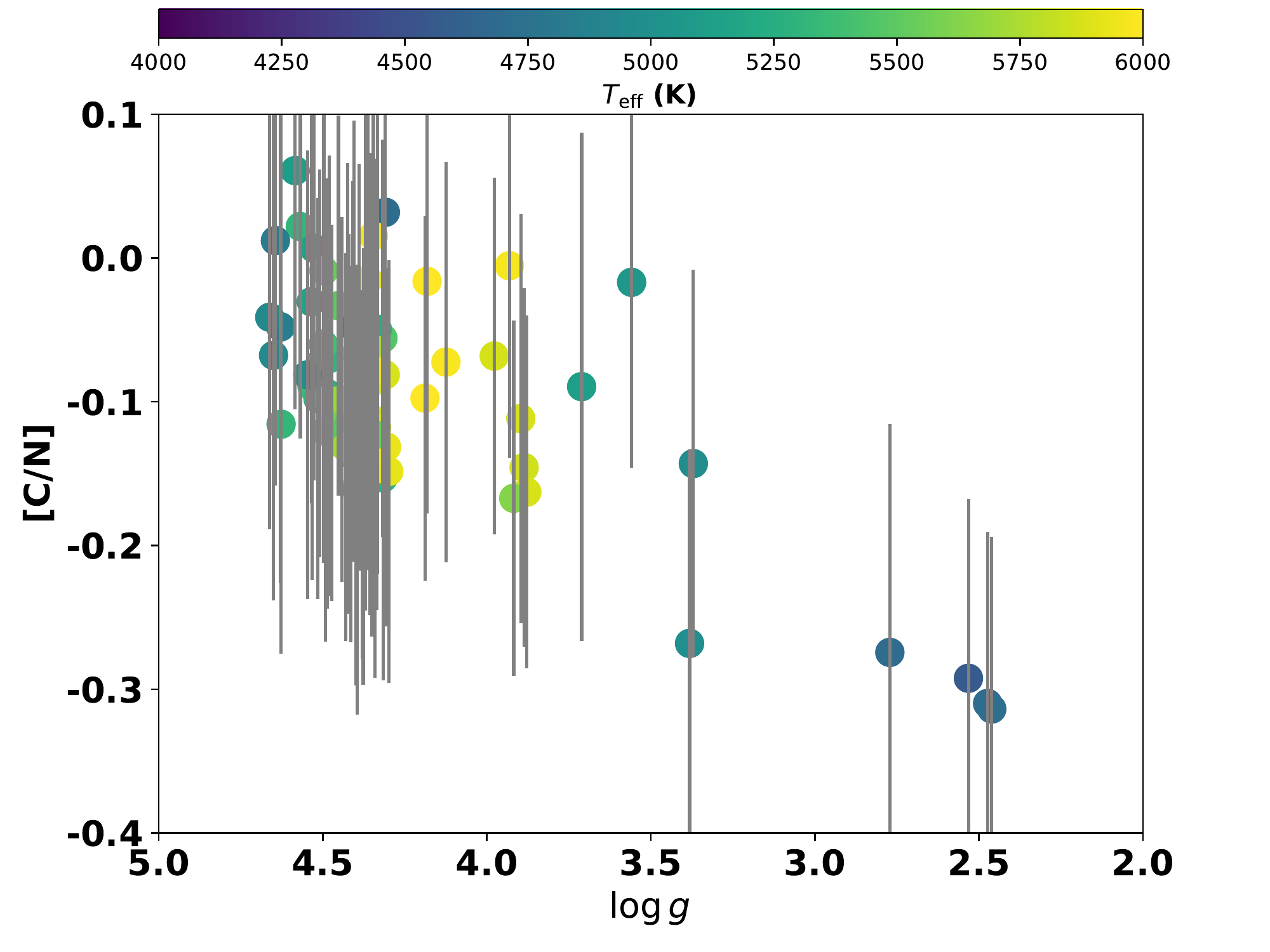}
\caption{[C/N] variations with $\log\,g$ of cluster members in the open cluster of  $\rm NGC\,2682$. [C/N] is equal to [C/Fe]$-$[N/Fe]. The gray lines show the error of each [C/N] value.}
\label{cluster_cn_giant}
\end{figure*}

 During red giant stage, the dredge-up process  change the  surface chemical abundance because of the convective mixing, which will finally decrease the surface abundance ratio of [C/N] \citep[e.g.,][]{Casali2019, Bertelli2017}.  As shown in  Fig.\,\ref{cluster_cn_giant}, we find a clear decrease of [C/N]  from $-0.05$ of dwarf stars to $-0.32$ of giant stars in the open cluster of $\rm NGC\,2682$ (M67), which is consistent with the results of \cite{Casali2019} and \cite{Bertelli2017}. Our estimated [C/Fe] and [N/Fe] could well reproduce the variations of [C/N] of red giant stars in open clusters.  [C/N] ratios  are tightly correlated with the stellar masses for red giants and thus can be further used to derive their ages using the relation of [C/N] and stellar masses \citep[e.g.,][]{Martig2016a, Ness2016}. One can use the estimated [C/Fe] and [N/Fe] values here of red giant stars to derive their stellar ages.

\subsection{The external comparisons of absolute magnitudes}
We test the quality of  our estimates of absolute magnitudes  through comparing with photometric absolute magnitudes.
Through cross-matching  the  LAMOST DR8 and Gaia\,EDR3, 2MASS, APASS, SDSS and WISE, we obtain 66,498 common stars. They have accurate estimations of distances, apparent magnitudes, interstellar extinctions and  high quality   LAMOST spectra (with spectral SNR\,$> 20$). For these stars we could estimate accurate photometric absolute magnitudes based on the distance modulus.   Then we  compare our estimated absolute magnitudes with the photometric  absolute magnitudes  in order to evaluate the quality of our estimates of absolute magnitudes.  

\begin{figure*}
\centering
\includegraphics[width=5.5in]{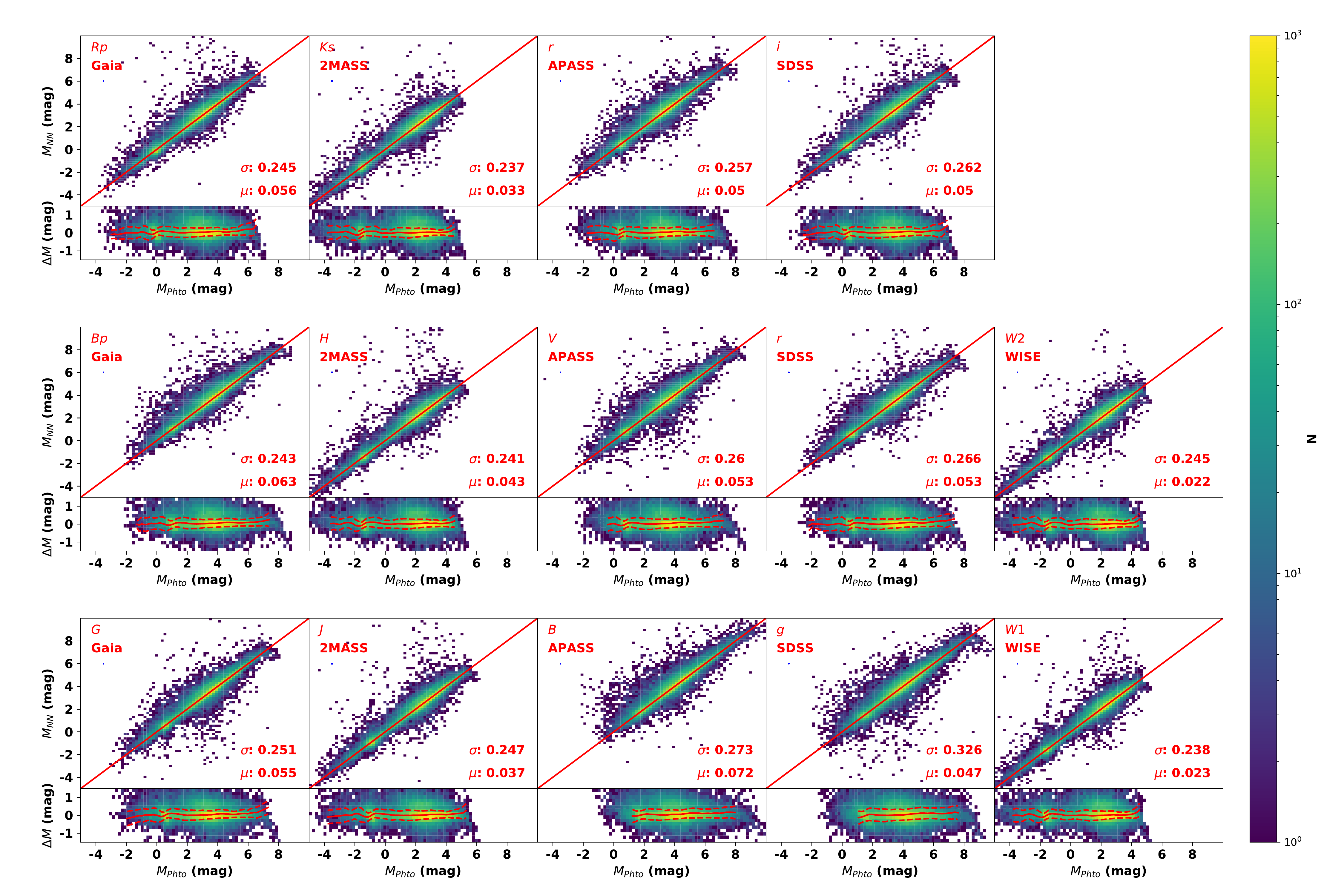}
\caption{Comparisons between the photometric absolute magnitudes estimated based on distance modulus  and absolute magnitudes derived from LAMOST spectra.  The offset and standard deviations  are marked in each panel. The difference are also overplotted at the bottom of each panel.   The absolute magnitudes based on distance modulus minus  that  derived from LAMOST spectra  are $\Delta M$. Typical errors of these absolute magnitudes are also indicated by the  error bars (blue cross) at the top left of each panel. However, they are hard to see because the typical errors are  very small. } 
\label{compare_photometric}
\end{figure*}

Fig.\,\ref{compare_photometric} shows the comparison results.   The two kinds of absolute magnitudes could match with each other very well.  
We find that the differences of the $M_{G},M_{Bp}, M_{Rp}, M_{J}, M_{H}, M_{Ks}, M_{B}$, $M_{V}, M_{g}, M_{r}, M_{i}, M_{r_{A}}, M_{W1}, M_{W2}$ between the two kinds estimates have  $\sigma$ of 0.251, 0.243, 0.245, 0.247, 0.241, 0.237, 0.273, 0.26, 0.257, 0.326, 0.266, 0.262, 0.238 and 0.245\,mag, respectively.  These $\sigma$ are equivalent to the distance error of $<$\,15\%.  The absolute value of systematic differences of absolute magnitudes of these bands are almost smaller than 0.1\,mag.

In conclusion, the stellar parameters estimated using neural network method have small uncertainties through comparing results derived from duplicate observations of similar spectral SNRs (differed by less than 10\%) collected during different nights. 
 For stars with spectral SNRs larger than 50,  precisions of  $T_{\mathrm{eff}}$,  $\log g$ (estimated using LAMOST-APOGEE common stars as training set),  [Fe/H] (estimated using LAMOST-APOGEE common stars as training set), [M/H], [C/Fe], [N/Fe] and [$\alpha$/M] are 85\,K, 0.098\,dex, 0.05\,dex, 0.05\,dex, 0.052\,dex, 0.082\,dex and 0.027\,dex, respectively.  The errors of 14 band’s absolute magnitudes are only 0.16--0.22\,mag for stars with spectral SNRs larger than 50.  As discussed in Section\,3, the $\log g$,  [Fe/H]  have two values for each star estimated using different stellar sample as training set (LAMOST-PASTEL common stars, LAMOST-APOGEE common stars).  Precisions of $\log g$  and [Fe/H] estimated using LAMOST-PASTEL common stars as training set are 0.135\,dex and 0.08\,dex for stars with spectral SNRs larger than 50. 
 Our stellar parameters could match well with other external parameter catalogs, including Gaia-ESO, APOGEE (for  $T_{\mathrm{eff}}$,  $\log g$,  [Fe/H], [M/H], [C/Fe], [N/Fe] and [$\alpha$/M] comparisons), catalog of asteroseismic measurements (for $\log g$ comparisons) and catalog of  photometric absolute magnitudes (for absolute magnitude comparisons).  For metallicity and chemical abundance ratios, we also studied the dispersions of stars in open  clusters. The dispersions are very small.  

\section{The [Fe/H] values of  very metal poor stars}
Very metal poor (VMP) stars ([Fe/H] $\leq -2.0$)  provide the fossil record of the early  chemical history of the Galaxy and early generations of stars.  In this section, we will present the improved estimates for VMP candidates using the neural network method. 

Firstly,  we select  65,465 stars with [Fe/H] $< -1.5$ and $T_{\rm eff} < 6500$\,K as VMP candidates and re-estimate their [Fe/H] values. In this process, we only use metal poor stars ([Fe/H] $<= -1.0$) in the pre-mentioned LAMOST--PASTEL training sample  as training stars.  Because the spectral features of VMPs are dominated in the blue range of LAMOST spectra, we only use  3900--4400\,\AA\,to determine the [Fe/H] values for metal poor stars.  Our final   neural network model contain two layers. The neurons of the first and second  layer of neural network model are respectively 256 and 64.   Using the neural network model based on the metal poor stars as training set, we estimate the [Fe/H] values for training stars using LAMOST spectra and compare with that provided by PASTEL.  The standard deviations of the residuals and systematic difference   are  only 0.176 and -0.01\,dex, respectively.  The improved  [Fe/H] values ([Fe/H]-NN-VMP) here could match well with that provided by PASTEL at $-1.0 <$ [Fe/H] $< -4.0$\,dex. One can see Fig.\,\ref{train_vmp} in Appendix for more details about the comparisons between the improved   [Fe/H] values and that provided by PASTEL.  After building up neural network model, we apply them to all LAMOST  low-resolution spectra of these VMP  candidates and estimate the improved  [Fe/H] values.

\cite{Li2018}  provide a catalog of 10,008 VMPs from LAMOST DR3, which is the largest number of VMP candidates.  The [Fe/H] values of these VMP stars are derived from LAMOST low resolution spectra.   70 of stars in  these stars have high-resolution spectroscopic analyses reported in the literature \citep{Fulbright2000, Carretta2002, Cohen2004, Cohen2006, Cohen2013, Honda2004, Aoki2005, Arnone2005, Barklem2005, Lai2007, Lai2008, Andrievsky2009, Andrievsky2010, Rich2009, Melendez2010, Caffau2011, Hollek2011, Suda2011, Allen2012, Bonifacio2012, Yong2013, Placco2014, Roederer2014, Placco2014, Li2015a, Li2015, Li2018, Jacobson2015, Matsuno2017}. Through comparing our un-improvded ([Fe/H]-NN-PASTEL) and  improved [Fe/H] values  with previous [Fe/H] estimates for VMP candidates,  the accuracy of  [Fe/H] values of our VMP candidates could be tested.  

\begin{figure*}
\centering
\includegraphics[width=5.5in]{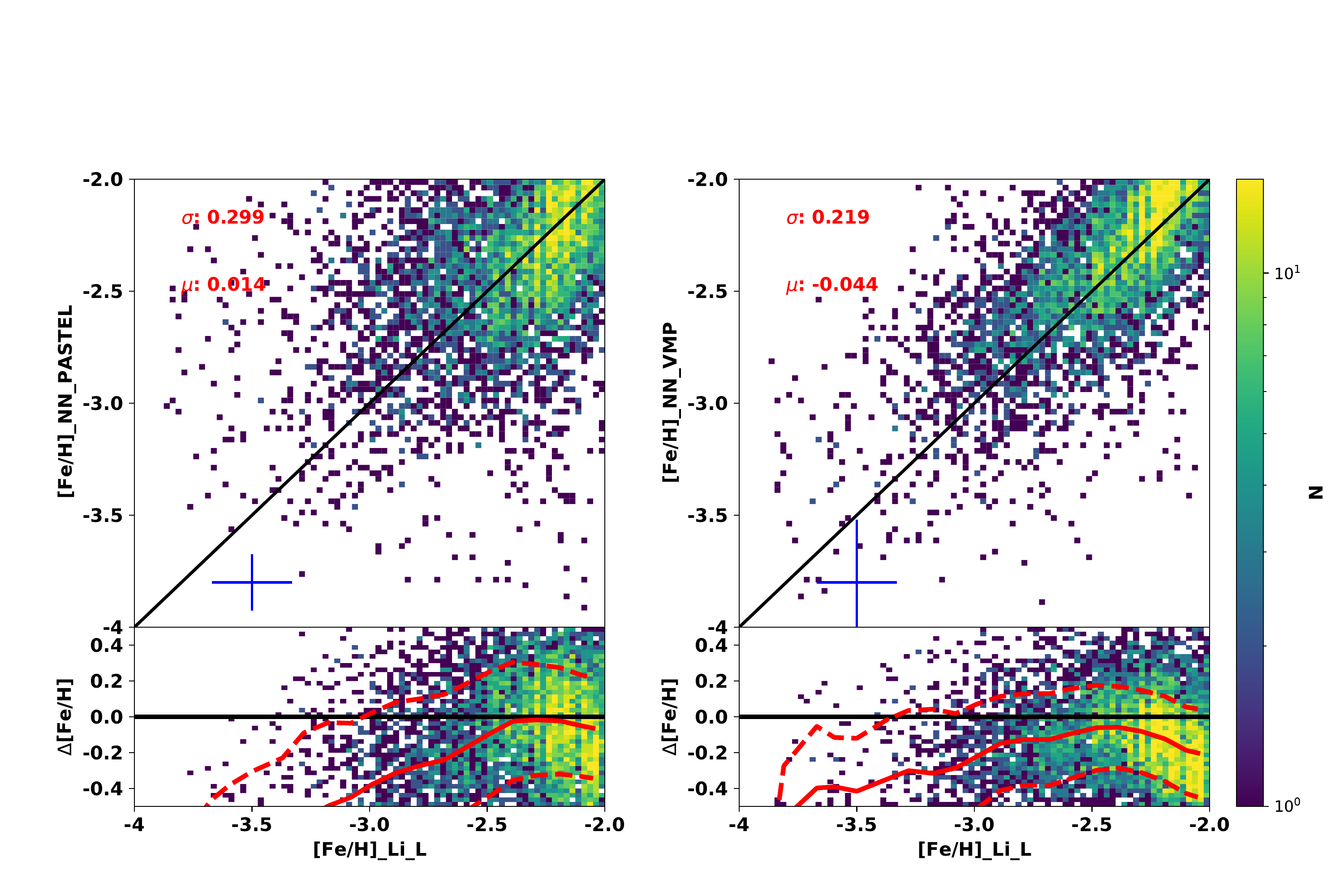}
\caption{The comparisons of  our [Fe/H]-NN-PASTEL (left panel) and [Fe/H]-NN-VMP (right panel) values with the [Fe/H] values provided by \cite{Li2018} for VMP  candidates.  Our [Fe/H]-NN-VMP and  [Fe/H]-NN-PASTEL minus [Fe/H]-Li-L are $\Delta \rm [Fe/H]$. Typical errors of [Fe/H] values of \cite{Li2018} and us are  indicated by the  error bars (blue cross) at the bottom left of each panel.}
\label{compare_vmp}
\end{figure*}

Fig.\,\ref{compare_vmp} shows the comparisons  between our two kinds of [Fe/H] values ([Fe/H]-NN-PASTEL and [Fe/H]-NN-VMP) and  [Fe/H] values (derived from LAMOST low resolution spectra: [Fe/H]-Li-L) provided by \cite{Li2018}.  
The  systematic difference between [Fe/H]-NN-PASTEL  and [Fe/H]-Li-L is 0.014\,dex, the sigma of the corresponding difference is 0.299\,dex. It is noted that the systematic differences between  the [Fe/H]-NN-PASTEL  and  [Fe/H]-Li-L vary with metallicity, [Fe/H]-NN-PASTEL are underestimated and overestimated for stars with high metallicity ($-2.3 < $[Fe/H]$ < -2.0$\,dex) and low metallicity ($-4.0 <$ [Fe/H] $< -2.3$\,dex), respectively. However, the   systematic differences  are almost smaller than 0.4\,dex for stars with [Fe/H] $ < -3.0$.   For the [Fe/H]-NN-VMP, the systematic difference and the sigma for the comparison with [Fe/H]-Li-L  are respectively only -0.044\,dex and 0.219\,dex. The [Fe/H]-NN-VMP are overestimated for stars with metallicity of $-4.0 <$ [Fe/H] $< -2.3$\,dex.  The overestimations of the [Fe/H]-NN-VMP is smaller than that of  [Fe/H]-NN-PASTEL. 

\begin{figure*}
\centering
\includegraphics[width=5.5in]{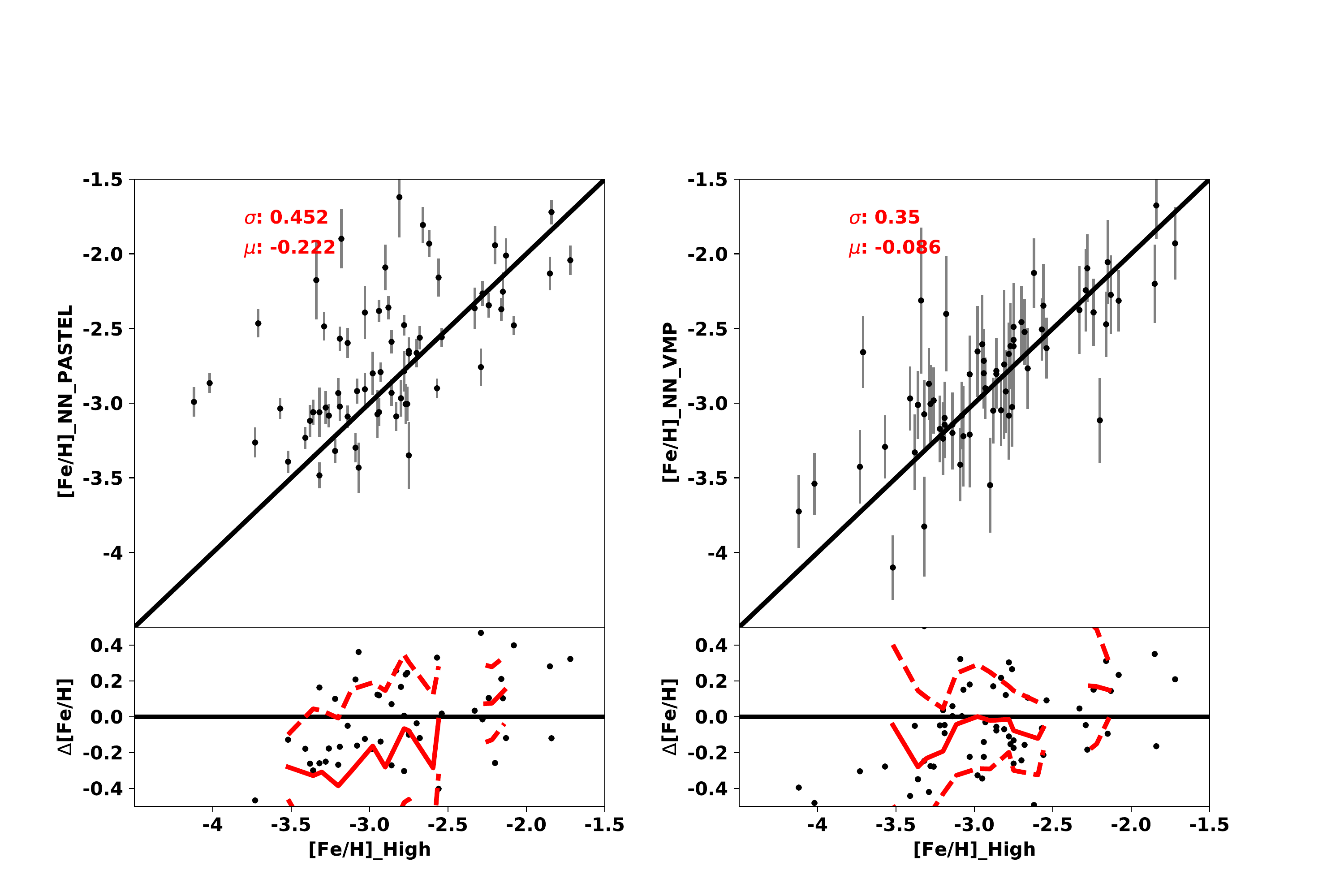}
\caption{The comparisons of  our [Fe/H]-NN-PASTEL (left panel) and [Fe/H]-NN-VMP (right panel) values with the [Fe/H] values derived from high resolution spectra.}
\label{compare_vmp_h}
\end{figure*}

Fig.\,\ref{compare_vmp_h} shows the comparison between  our two kinds of [Fe/H] values and   [Fe/H] values analysed based on high-resolution spectra.   We find that the difference with the  [Fe/H] values of high-resolution of [Fe/H]-NN-PASTEL and [Fe/H]-NN-VMP have sigma of  0.452\,dex and 0.35\,dex, respectively.  The average measurement offset of [Fe/H]-NN-PASTEL and [Fe/H]-NN-VMP are respectively -0.222\,dex and -0.086\,dex.    The [Fe/H]-NN-VMP could match well with that derived from high resolution spectra for stars with$-4.0 < $ [Fe/H] $\leq -1.5$\,dex.  

Through comparing with [Fe/H]-Li-L and [Fe/H] values of high-resolution, we suggest that  the  [Fe/H]-NN-VMP are much better than the [Fe/H]-NN-PASTEL for stars with [Fe/H] $< -1.5$\,dex.  Based on the  [Fe/H]-NN-VMP values,   we select 26,868 unique stars with SNR larger than 20 as VMP candidates ([Fe/H] $\leq -2.0$\,dex). Among of them,  3,952 unique stars  have  [Fe/H] $\leq -3.0$\,dex. Our catalog will provide the largest number of VMP candidates up to now.




\section{Distance}

Now many distances of stars in the Milky Way could be derived using parallax provided by Gaia EDR3. The errors of parallaxes provided by Gaia EDR3 depend on the distance of stars.  For distances larger than $\sim$\,2.0\,kpc, the  systematic errors of parallax-based distance estimates become significant \citep{Lindegren2018, Huang2021a, Lindegren2021, Lindegren2021b, Ren2021, Xiang2021}.  For stars with distances larger than $\sim$\,2.0\,kpc, spectro-photometric distance estimation is a better choice. The accuracy of spectro-photometric distance  does not depend on itself.  

We provide spectro-photometric distance based on absolute magnitudes directly derived from the LAMOST spectra. 
Distance of  individual stars observed by LAMOST are estimated with distance modulus method  using the   apparent magnitudes, interstellar extinctions and our absolute magnitudes.  Here, we adopt $M_{Ks}$  to calculate  distance,  as it suffers from less severe extinction than other bands. After estimating distance, we estimate the distance  errors based on  the principle of uncertainty propagation.  In doing so, errors of absolute magnitudes, extinction and apparent magnitudes are considered.  The spectro-photometric distances are accurate  to 8.5\% for stars with spectral SNRs larger than 50. 

\begin{figure}
\centering
\includegraphics[width=3.5in]{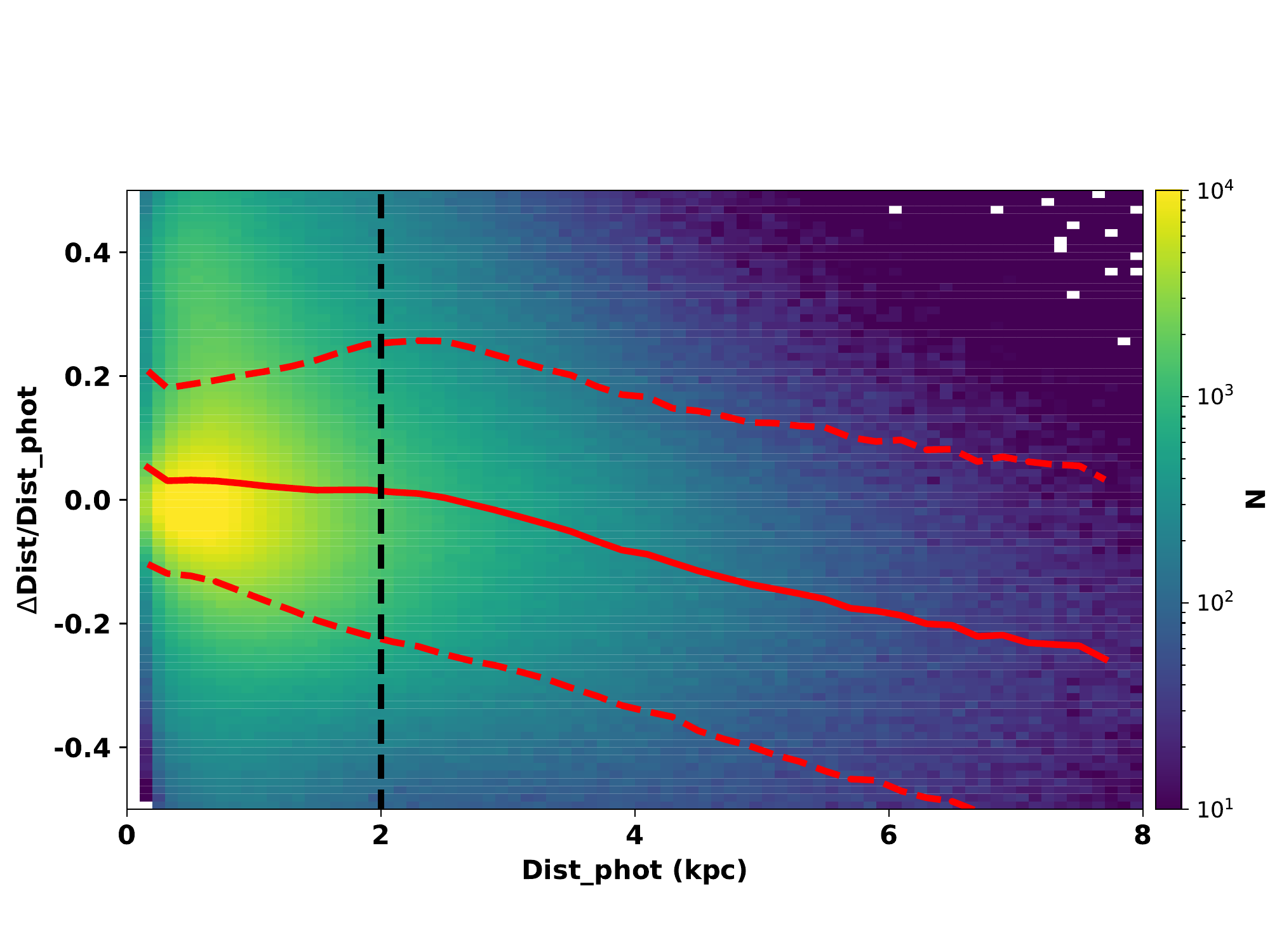}
\caption{ The variations of systematic difference  of our spectro-photometric distance and geometrical distance provided by Gaia\,EDR3 with our spectro-photometric distance.  The geometrical distance  minus   our spectro-photometric distance  are $\Delta Dist$. The black dashed line show the spectro-photometric distance of 2.0\,kpc.}
\label{compare_dist}
\end{figure}

Fig.\,\ref{compare_dist} shows the variations of systematic difference  of our spectro-photometric distance and geometrical distance provided by Gaia\,EDR3 with our spectro-photometric distance.  We can find that systematic difference  is $\sim$ 0 at $D \leq 2.0$\,kpc, which suggest that  geometrical distance could match well with  spectro-photometric distance for stars with $D \leq 2.0$\,kpc. However, for stars with distance larger than 2.0\,kpc, the systematic difference become larger  with distance increasing.  Up to 8\,kpc, the geometrical distances are underestimated  about 25\,per\,cent, which is similar with that of \cite{Xiang2021}.

\section{The value-added catalog}
After estimating aforementioned stellar parameters, we build up a value-added catalog for LAMOST\,DR8 low-resolution stellar spectra. The value-added catalog contains the stellar atmospheric parameters ($T_{\rm eff}$, $\log\,g$ and [Fe/H]/[M/H]), chemical abundance ratios ([$\alpha$/M], [C/Fe] and [N/Fe]), 14 band's absolute magnitudes ($M_{G},M_{Bp}, M_{Rp}$ of Gaia EDR3 bands, $M_{J}, M_{H}, M_{Ks}$ of 2MASS bands, $M_{W1}, M_{W2}$ of WISE bands, $M_{B}, M_{V}, M_{r_{A}}$ of APASS bands and $M_{g}, M_{r}, M_{i}$ of SDSS bands) and spectro-photometric distances of 7,109,030 spectra of 5,166,619  unique stars. The value-added catalog contains stellar parameters for all LAMOST spectra with SNRs larger than 10.  Because of  bad LAMOST spectra, the effective stellar parameter range of training sets as discussed in section\,3 and systematic errors of stellar parameters for some stars as discussed in section\,4, some stellar parameter estimates are not so accurate. One need make some selection criteria of stellar parameters according to their science goals.
Extinctions from Huang et al. (2021) are also given. 
The catalog  also provide the radial velocities derived by LASP. 
A detailed description of the catalog content is presented in Table\,\ref{table3}.  The catalog will be released in the website of \url{http://www.lamost.org/dr8/v1.0/doc/vac}.

\begin{figure*}
\centering
\includegraphics[width=5.5in]{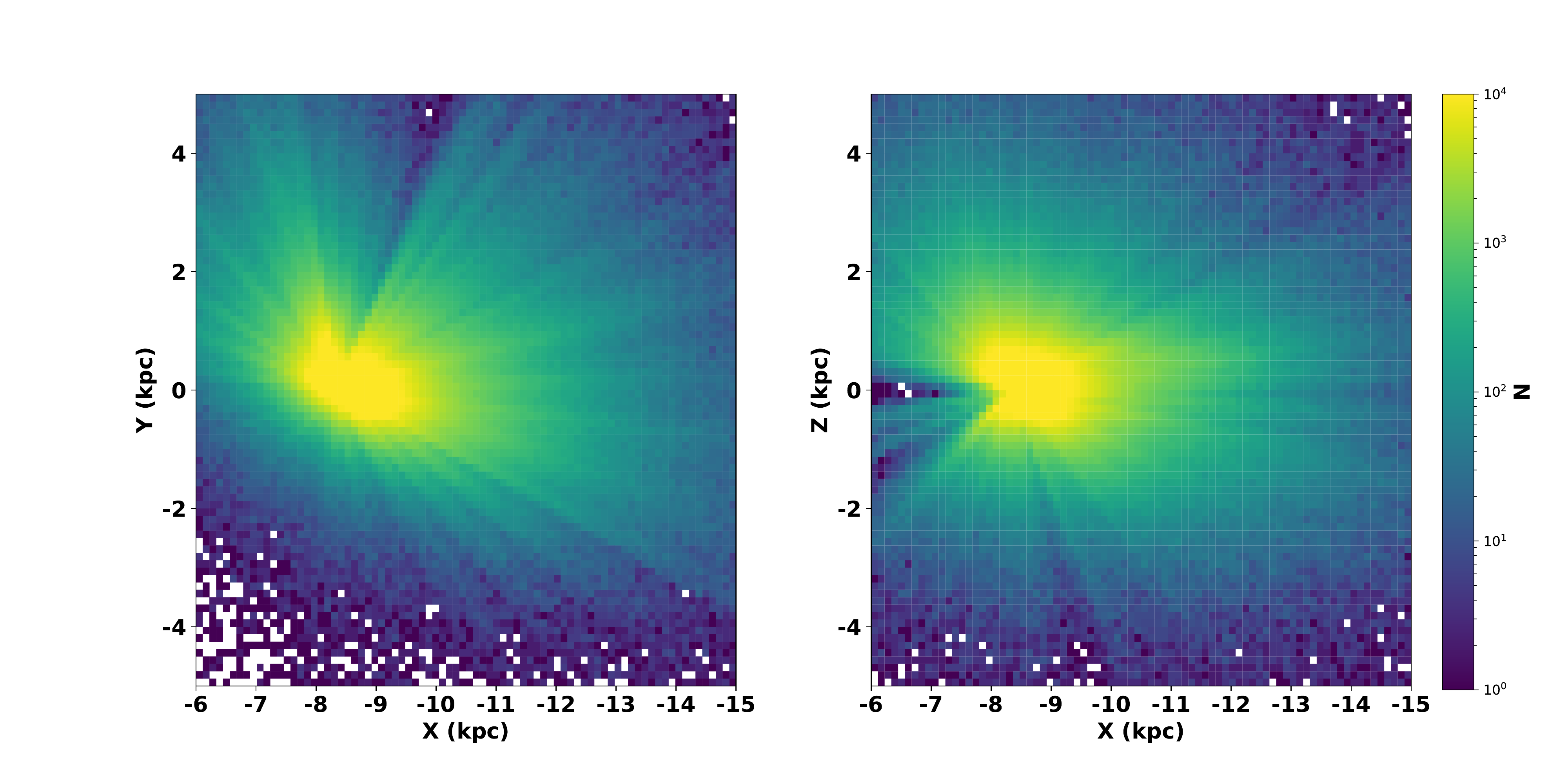}
\caption{ Spatial stellar number density distributions of the LAMOST\,DR8 stars in the disc $X$--$Y$ (left panel) and $X$--$Z$ (right panel)
planes. }
\label{xyxz}
\end{figure*}

\begin{figure*}
\centering
\includegraphics[width=5.5in]{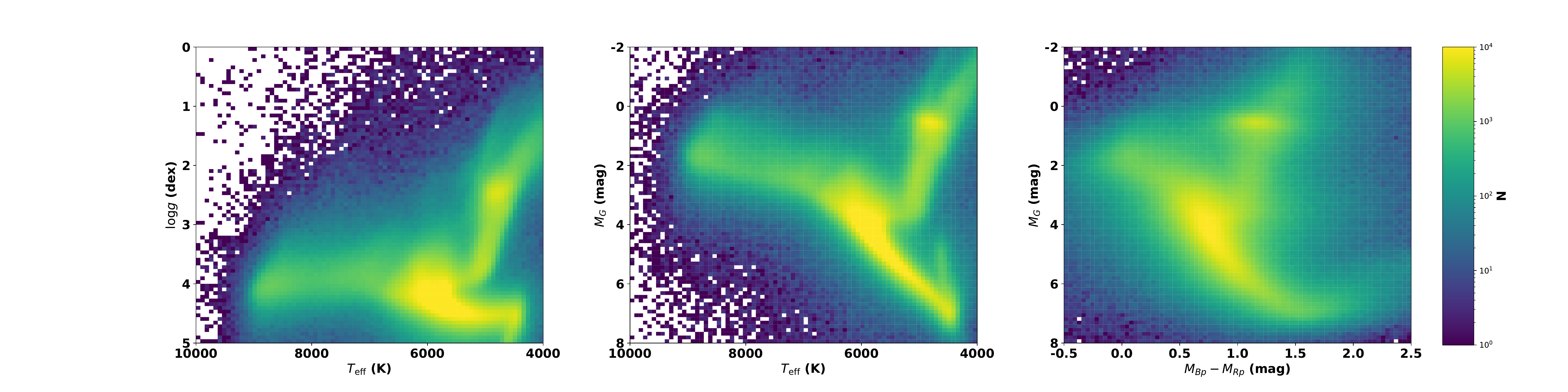}
\caption{ The stellar number density distributions of the LAMOST\,DR8 stars in the $T_{\rm eff}$--$\log g$ (left panel), $T_{\rm eff}$--$M_{G}$ (medium panel) and ($M_{Bp}-M_{Rp}$)--$M_{G}$ (right panel)  planes. }
\label{hr_diagram}
\end{figure*}

\begin{figure*}
\centering
\includegraphics[width=5.5in]{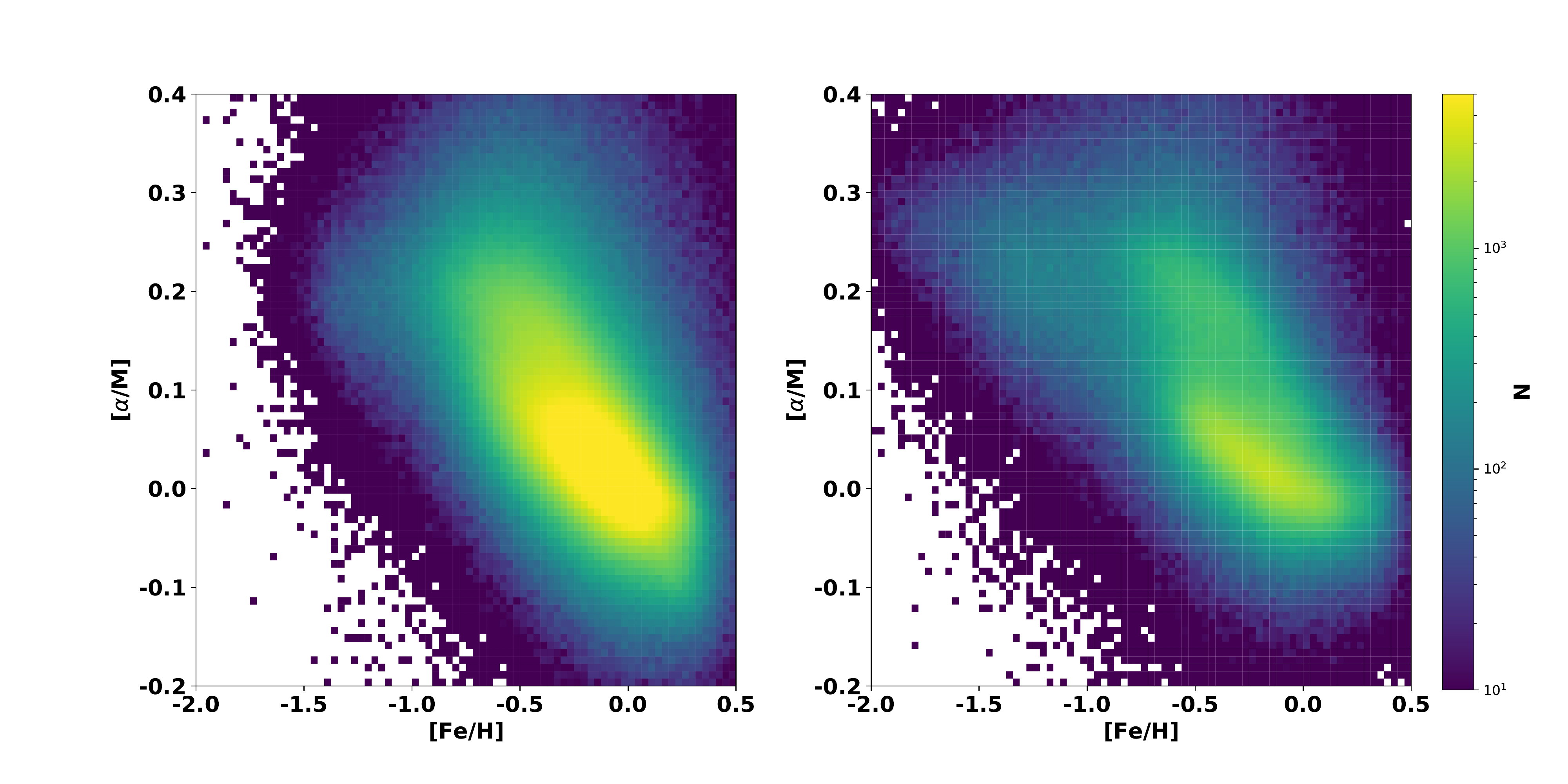}
\caption{ The stellar number density distributions of the LAMOST\,DR8 stars in the [Fe/H]-[$\alpha$/M] plane for dwarf (left panel) and giant (right panel) stars.  }
\label{fehafe}
\end{figure*}

We estimate 3D positions in the right-handed Cartesian system ($X$, $Y$, $Z$ with $X$ toward the direction opposite to the Sun, $Y$ in the direction of Galactic rotation, and $Z$ toward the north Galactic pole).  Fig.\,\ref{xyxz} shows the spatial stellar number density distributions of our stellar sample  in the $X$--$Y$ and $X$--$Z$ planes. This figure suggests that our sample covers a  large volume of the Milky Way.  One can use it to probe  the structural, kinematic and chemical  properties of the Galactic disc combining the proper motions provided by Gaia\,EDR3.  Fig.\,\ref{hr_diagram} shows the stellar number density distributions of stars in the $T_{\rm eff}$--$\log g$, $T_{\rm eff}$--$M_{G}$ and ($M_{Bp}-M_{Rp}$)--$M_{G}$  planes. 
Fig.\,\ref{fehafe} shows the stellar number density distributions of our stellar sample in the [Fe/H]-[$\alpha$/M] plane for dwarfs and giants.  The two figures suggest that the stellar parameters of our sample span wide ranges, contain numerous dwarf stars, main-sequence turn-off stars, giant stars, metal-poor and metal-rich stars. One could build up different stellar sample for different scientific goals using this value-added catalog. 

\clearpage
\begin{longtable}{ccc}
\caption{Descriptions for the value-added catalog.}\\
\label{table3}\\

\hline \hline
Field & Description & Unit \\
\endfirsthead
\hline
gaiaid&Cross-matched Gaia EDR3 source ID &-- \\
obsid&Unique number ID of this spectrum &-- \\
ra\_obs&Right ascension from the LAMOST (J2000) &degree \\
dec\_obs&Decl. from the LAMOST (J2000) &degree \\
gl&Galactic longitude derived from ICRS coordinates &degree \\
gb&Galactic latitude derived from ICRS coordinates &degree \\
snrg&The LAMOST spectral SNRs in $g$ band&1 \\
rv&Radial velocity&$\rm km\,s^{-1}$\\
Err\_rv&Error of radial velocity&$\rm km\,s^{-1}$\\
$T_{\rm eff}$\_PASTEL&Effective temperature estimated with neural network using LAMOST-PASTEL sample as training set &K \\
Err\_$T_{\rm eff}$\_PASTEL&Errors of  $T_{\rm eff}$\_PASTEL&K \\
$\log\,g$\_PASTEL & Surface gravity estimated with neural network using LAMOST-PASTEL sample as training set &  dex  \\
Err\_$\log\,g$\_PASTEL & Errors of  $\log\,g$\_PASTEL & dex \\
$\log\,g$\_APOGEE & Surface gravity estimated with neural network using LAMOST-APOGEE sample as training set &  dex  \\
Err\_$\log\,g$\_APOGEE & Errors of  $\log\,g$\_APOGEE& dex \\
 feh\_PASTEL&The value of [Fe/H] with neural network using LAMOST-PASTEL sample as training set&dex \\
Err\_feh\_PASTEL& Errors of  feh\_PASTEL & dex \\
feh\_APOGEE&The value of [Fe/H] with neural network using LAMOST-APOGEE sample as training set&dex \\
Err\_feh\_APOGEE& Errors of  feh\_APOGEE & dex \\
feh\_VMP&The improved [Fe/H] values of VMP candidates&dex \\
Err\_feh\_VMP&Errors of  feh\_VMP&dex \\
MH\_APOGEE&The value of [M/H] with neural network using LAMOST-APOGEE sample as training set&dex \\
Err\_MH\_APOGEE& Errors of  the [M/H]&dex\\
AFE\_APOGEE&The value of [$\alpha$/M] with neural network using LAMOST-APOGEE sample as training set&dex \\
Err\_AFE\_APOGEE& Errors of  the  [$\alpha$/M] &dex\\
CFE\_APOGEE&The value of [C/Fe] with neural network using LAMOST-APOGEE sample as training set&dex \\
Err\_CFE\_APOGEE& Errors of  the  [C/Fe] &dex\\
NFE\_APOGEE&The value of [N/Fe] with neural network using LAMOST-APOGEE sample as training set&dex \\
Err\_NFE\_APOGEE& Errors of  the [N/Fe]&dex\\
dist\_phot& Distances estimated based on distance modulus & kpc\\
Err\_dist\_phot&Errors of distances& kpc\\
a\_gg &Absolute magnitude of $Gaia$\,EDR3 $G$ band &mag\\
Err\_a\_gg &Error of absolute magnitude of $Gaia$\,EDR3 $G$ band&mag\\
a\_bp&Absolute magnitude of $Gaia$\,EDR3 $Bp$ band&mag\\
Err\_a\_bp&Error of absolute magnitude of $Gaia$\,EDR3 $Bp$ band&mag\\
a\_rp&Absolute magnitude of $Gaia$\,EDR3 $Rp$ band&mag\\
Err\_a\_rp&Error of absolute magnitude of $Gaia$\,EDR3 $Rp$ band&mag\\
a\_j&Absolute magnitude of 2MASS $J$ band&mag\\
Err\_a\_j&Error of absolute magnitude of 2MASS $J$ band&mag\\
a\_h&Absolute magnitude of 2MASS $H$ band&mag\\
Err\_a\_h&Error of absolute magnitude of 2MASS $H$ band&mag\\
a\_ks&Absolute magnitude of 2MASS $Ks$ band&mag\\
Err\_a\_ks&Error of absolute magnitude of 2MASS $Ks$ band&mag\\
a\_w1&Absolute magnitude of WISE $W1$ band&mag\\
Err\_a\_w1&Error of absolute magnitude of WISE $W1$ band&mag\\
a\_w2&Absolute magnitude of WISE $W2$ band&mag\\ 
\hline \hline
Field & Description & Unit \\
\hline

Err\_a\_w2&Error of absolute magnitude of WISE $W2$ band&mag\\
a\_bap&Absolute magnitude of APASS $B$ band&mag\\
Err\_a\_bap&Error of absolute magnitude of APASS $B$ band&mag\\
a\_rap&Absolute magnitude of APASS $r$ band&mag\\
Err\_a\_rap&Error of absolute magnitude of APASS $r$ band&mag\\
a\_vap&Absolute magnitude of APASS $V$ band&mag\\
Err\_a\_vap&Error of absolute magnitude of APASS $V$ band&mag\\
a\_gsd&Absolute magnitude of SDSS $g$ band&mag\\
Err\_a\_gsd&Error of absolute magnitude of SDSS $g$ band&mag\\
a\_rsd&Absolute magnitude of SDSS $r$ band&mag\\
Err\_a\_rsd&Error of absolute magnitude of SDSS $r$ band&mag\\
a\_isd&Absolute magnitude of SDSS $i$ band&mag\\
Err\_a\_isd&Error of absolute magnitude of SDSS $i$ band&mag\\
X/Y/Z&3D positions in the right-handed Cartesian system&kpc\\
Err\_X/Y/Z&Errors of 3D positions in the right-handed Cartesian system&kpc\\
EBV&interstellar extinctions estimated using ``star-pair" method&--\\
uqflag&Uniqueness flag. 1 denoting the spectrum of the highest snr\_g of one star&--\\
flag\_parameter& The quality   flag$^{\rm a}$ on every parameter &unit\\
\colrule

\end{longtable}
\footnotesize{$^a$ } The flag takes the values of 0 and 1. Flag = 0 means that the parameter estimates 
are within  the training grid, the estimates are reliable.   Flag =1 means that the parameter estimates 
are beyond the training grid, the estimates are unreliable. \\
\clearpage

\section{Summary}
In the current work, we build up a value-added catalog for over 7.10 million LAMOST\,DR8 low-resolution stellar spectra with spectral SNR\,$>$\,10 of $5.16$ million unique stars.  We obtain the stellar atmospheric parameters ($T_{\rm eff}$, $\log\,g$ and [Fe/H]/[M/H]), chemical element abundances ([$\alpha$/M], [C/Fe] and [N/Fe]), 14 bands' absolute magnitudes ($M_{G},M_{Bp}, M_{Rp}$ of Gaia\,EDR3 bands, $M_{J}, M_{H}, M_{Ks}$ of 2MASS bands, $M_{W1}, M_{W2}$ of WISE bands, $M_{B}, M_{V}, M_{r_{A}}$ of APASS bands and $M_{g}, M_{r}, M_{i}$ of SDSS bands) and spectro-photometric distances using LAMOST spectra.  

Based on neural network method  adopting LAMOST-PASTEL sample as training set, we have estimated the $T_{\rm eff}$, $\log g$ and [Fe/H]. The accuracy of $T_{\rm eff}$, $\log g$ and [Fe/H] are respectively 84.87\,K, 0.135\,dex and 0.08\,dex for stars with LAMOST spectral SNR\,$> 50$.  Based on neural network method adopting LAMOST-APOGEE sample as training set, we have derived [Fe/H], [M/H], [$\alpha$/M], [C/Fe], [N/Fe] and $\log g$. The accuracy of [Fe/H], [M/H], [$\alpha$/M], [C/Fe], [N/Fe] and $\log g$  are respectively 0.05\,dex, 0.05\,dex, 0.027\,dex, 0.052\,dex, 0.082\,dex and 0.098\,dex for stars with LAMOST spectral SNR\,$> 50$. Based on neural network method adopting LGMWAS sample as training set, the  14 bands' absolute magnitudes have been derived. The uncertainties of $M_{G},M_{Bp}, M_{Rp}$ of Gaia EDR3 bands, $M_{J}, M_{H}, M_{Ks}$ of 2MASS bands,  $M_{B}, M_{V}, M_{r_{A}}$ of APASS bands, $M_{g}, M_{r}, M_{i}$ of SDSS bands and $M_{W1}, M_{W2}$ of WISE bands  are respectively 0.176, 0.163, 0.177, 0.186, 0.173, 0.177, 0.187, 0.174, 0.179,  0.221, 0.180, 0.184, 0.185 and 0.190\,mag for stars with LAMOST spectral SNR\,$> 50$.  
Using our accurate absolute magnitude of $M_{Ks}$, interstellar extinction and apparent magnitudes, we have derived the spectro-photometric distances based on the method of distance modulus. The accuracy of spectro-photometric distance  is $\sim$ 8.5\% for stars with LAMOST spectral SNR\,$> 50$. 

Benefiting from the good estimation of [Fe/H] values for metal poor stars, our catalog provide a largest number of VMP candidates (26,868 unique stars with [Fe/H] $\leq -2.0$\,dex) up to know.  Among of them,  3,952 unique stars have [Fe/H]  $\leq -3.0$\,dex.   These VMP candidates will play important roles in the study of  searching for  the first stellar populations and the study of the Milky Way halo.  For stars with $D > 2.0$\,kpc, our spectro-photometric distance are better than geometrical distance.  
 
The final sample  contain a large number of stars, cover a large and contiguous sky coverage and has simple yet non-trivial target selection function.  One can use it to probe  the structural, kinematic and chemical properties of the MW combining the proper motions provided by Gaia\,EDR3.

\begin{acknowledgments}
This work was funded by the National Key R\&D Program of China
(No. 2019YFA0405500) and the National Natural Science Foundation of China (NSFC Grant No.11833006, U1531244 and 11973001,  12173007).
Guoshoujing Telescope (the Large Sky Area Multi-Object Fiber Spectroscopic
Telescope LAMOST) is a National Major Scientific Project built by the Chinese Academy of Sciences.
Funding for the project has been provided by the National Development and Reform Commission.
LAMOST is operated and managed by the National Astronomical Observatories, Chinese Academy of
Sciences. The LAMOST FELLOWSHIP is supported by Special Funding for Advanced Users, budgeted and administrated by Center for Astronomical Mega-Science, Chinese Academy of Sciences (CAMS). Supported by High-performance Computing Platform of Peking University. 
\end{acknowledgments}



\bibliography{vac}{}
\bibliographystyle{aasjournal}

\appendix
  \renewcommand{\appendixname}{Appendix~\Alph{section}}
\setcounter{figure}{0}
\renewcommand{\thefigure}{A\arabic{figure}}

\begin{figure*}
\centering
\includegraphics[width=5.5in]{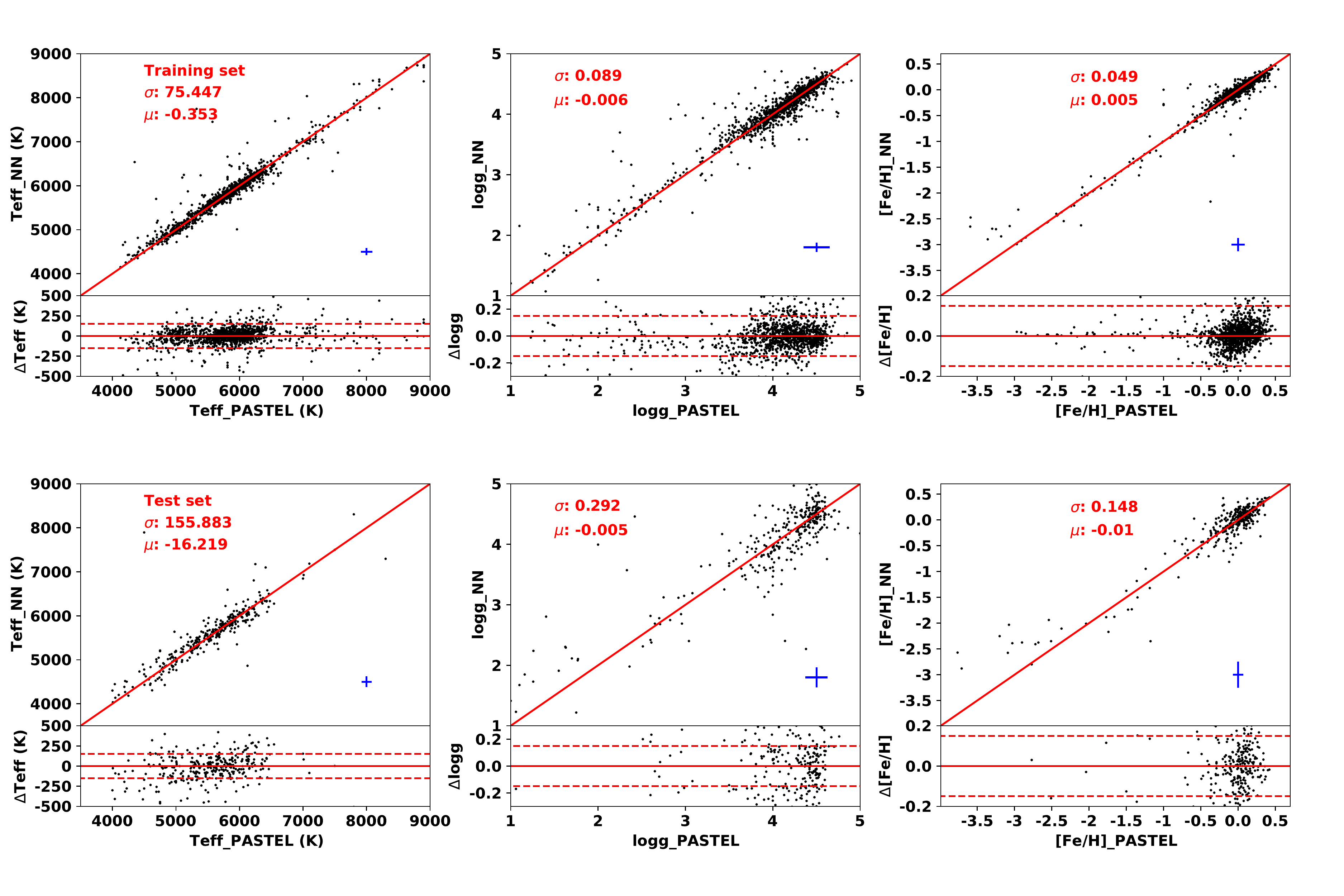}
\caption{ Comparisons between the stellar atmospheric parameters coming from  PASTEL and that derived from LAMOST spectra for training (top panels) and test (bottom panels) stars based on neural-network models.  The offset and standard deviations  are marked in  each panel. The difference are also overplotted at the bottom of each panel.  The values of stellar atmospheric parameters coming from PASTEL minus that  derived from LAMOST spectra are $\Delta T_{\rm eff}$, $\Delta\log g$ or $\Delta$[Fe/H]. Typical errors of $T_{\rm eff}$,  $ \log g$ and [Fe/H]  are  indicated by the  error bars (blue cross) at the bottom right of each panel.}
\label{pastel_test}
\end{figure*}

\begin{figure*}
\centering
\includegraphics[width=5.5in]{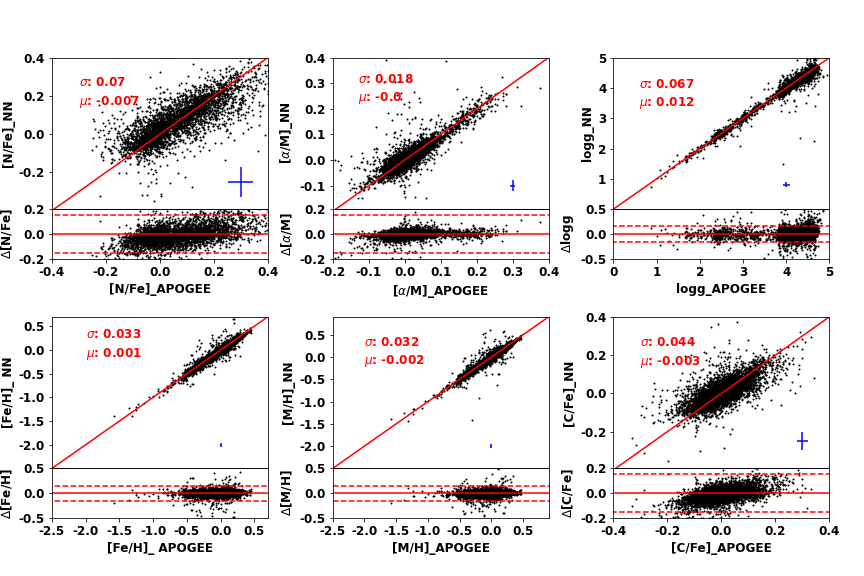}
\caption{Comparisons between the stellar parameters coming from  APOGEE and that derived from LAMOST spectra for LAMOST-APOGEE test stars based on neural-network models.  The offset and standard deviations  are marked in  each panel. The difference are also overplotted at the bottom of each panel. The values of stellar parameters coming from APOGEE minus that  derived from LAMOST spectra are $\Delta$[X/H], $\Delta$[X/Fe] or $\Delta\log g$.  The horizontal dashed lines indicate zero minus/plus 0.15\,dex. Typical errors of chemical abundances and $\log g$ are  indicated by the  error bars (blue cross) at the bottom right of each panel.}
\label{test_apogee}
\end{figure*}

\begin{figure*}
\centering
\includegraphics[width=5.5in]{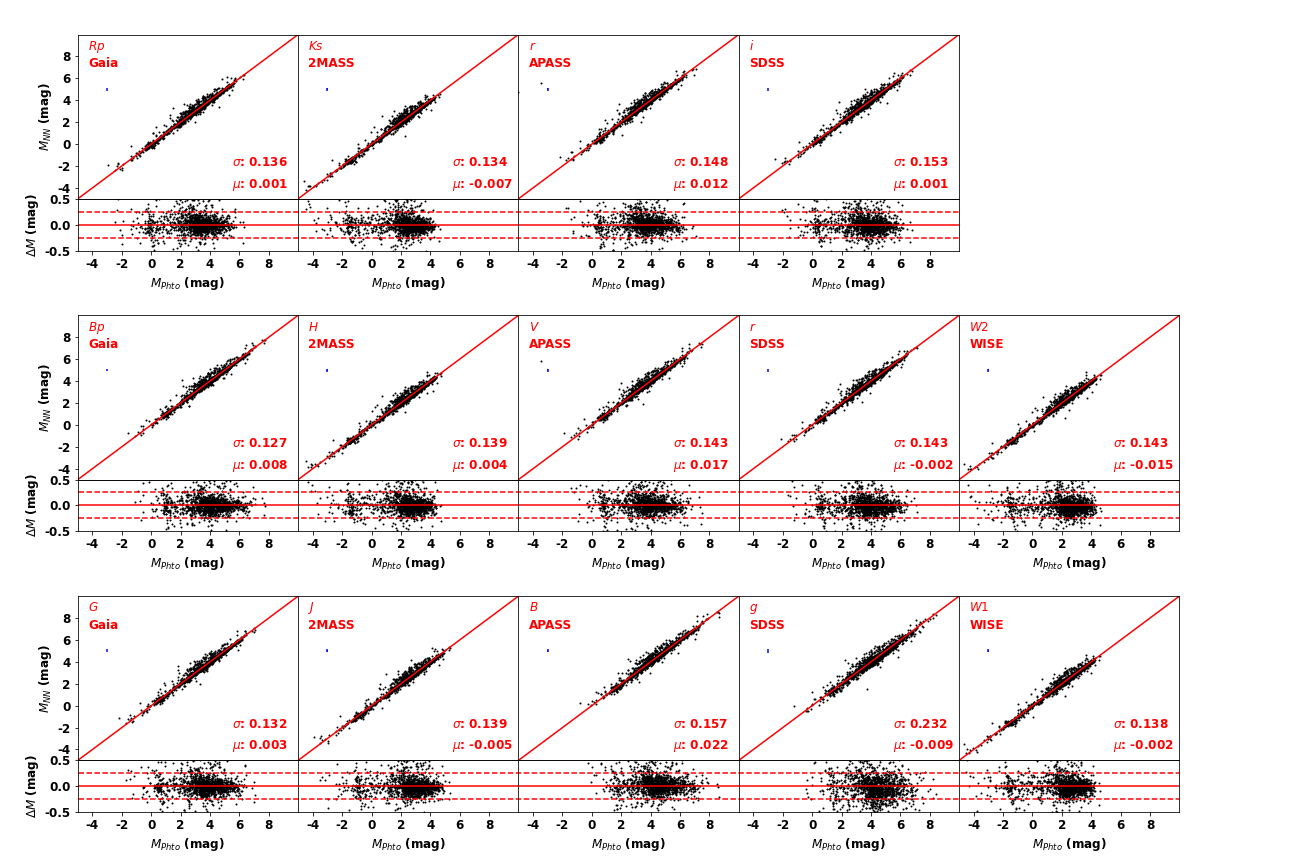}
\caption{Comparisons between the photometric absolute magnitudes estimated based on the distance modulus  and absolute magnitudes derived from LAMOST spectra for LGMWAS test  stars based on neural-network models.  The offset and standard deviations  are marked in  each panel. The difference are also overplotted at the bottom of each panel.   The absolute magnitudes based on distance modulus minus  that  derived from LAMOST spectra  are $\Delta M$. The horizontal dashed lines indicate zero minus/plus 0.25\,mag. Typical errors of these absolute magnitudes are also indicated by the error bars (blue cross) at the top left of each panel. However, they are hard to see because the typical errors are very small.}
\label{test_phot}
\end{figure*}

\begin{figure}
\centering
\includegraphics[width=3.5in]{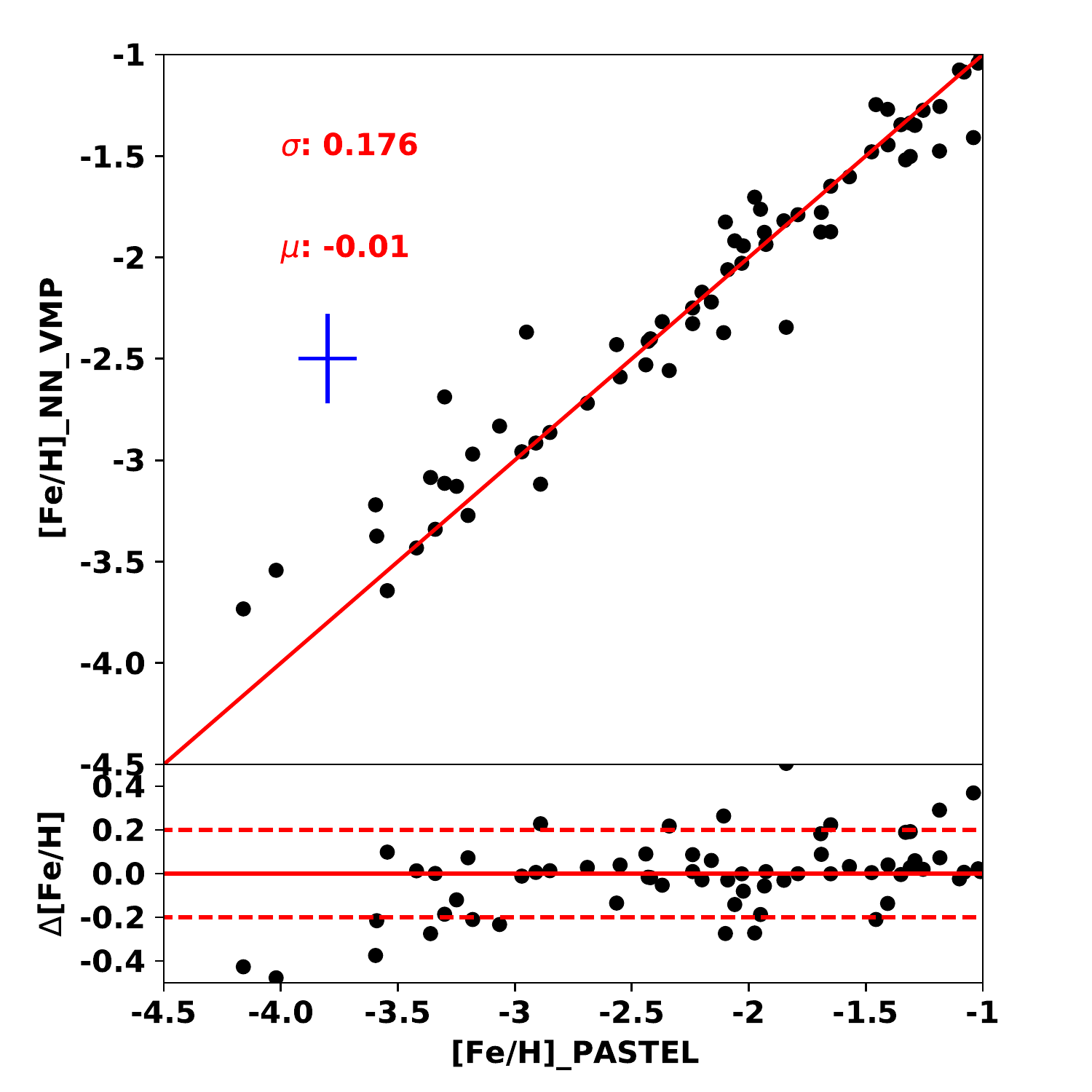}
\caption{Comparison between the [Fe/H] values coming from  PASTEL and our improved [Fe/H] values for LAMOST-PASTEL  [Fe/H] $<-1.0$\,dex training stars.  The offset and standard deviation  are marked in the figure. The difference are also overplotted at the bottom of the figure. The [Fe/H] values coming from PASTEL minus that  derived from LAMOST spectra are $\Delta$[Fe/H].  The horizontal dashed lines indicate zero minus/plus 0.2\,dex. Typical errors of [Fe/H] of us and PASTEL are  indicated by the  error bars (blue cross) at the top left.}
\label{train_vmp}
\end{figure}



\end{document}